\documentclass[journal=jctcce,manuscript=article]{achemso}
\usepackage{epsfig,graphicx}
\usepackage{amssymb,amsfonts,amsmath,bm}
\usepackage{hyperref}

\setkeys{acs}{usetitle = true}

\author{Sergei V. Krivov}
\affiliation[University of Leeds]
{Astbury Center for Structural Molecular Biology, Faculty of Biological Sciences, University of Leeds, Leeds LS2 9JT, United Kingdom}
\email{s.krivov@leeds.ac.uk}

\title{Blind analysis of molecular dynamics.}

\begin{document}
\maketitle

\begin{abstract}
We describe a non-parametric approach for accurate determination of the slowest relaxation eigenvectors of molecular dynamics. The approach is blind as it uses no system specific information. In particular, it does not require a functional form with many parameters to closely approximate eigenvectors, e.g., a linear combinations of molecular descriptors or a deep neural network, and thus no extensive expertise with the system. We suggest a rigorous and sensitive validation/optimality criterion for an eigenvector. The criterion uses only eigenvector timeseries and can be used to validate eignevectors computed by other approaches. The power of the approach is illustrated on long atomistic protein folding trajectories. The determined eigenvectors pass the validation test at timescale of $0.2$ ns, much shorter than alternative approaches. 
\end{abstract}

\section{Introduction}Molecular dynamics simulations increasingly produce massive trajectories \cite{shaw_atomic-level_2010, lindorff-larsen_how_2011}. Accurate analysis and interpretation of such data are widely recognized as fundamental bottlenecks that could limit their applications, especially in the forthcoming era of exascale computing \cite{freddolino_challenges_2010, schwantes_modeling_2015, banushkina_optimal_2016, noe_collective_2017, jung_artificial_2019,peters_common_2015,peters_reaction_2016}. A rigorous way to analyze dynamics in such data is to describe/approximate it by diffusion on the free energy landscape, free energy as a function of reaction coordinates (RCs). For such a description to be quantitatively accurate, the RCs should be chosen in an optimal way \cite{banushkina_optimal_2016, krivov_protein_2018}. The committor function is an example of such RCs, that can be used to compute some important properties of the dynamics exactly \cite{krivov_protein_2018}. The eigenvectors (EVs) of the transfer operator are another example \cite{shuler_relaxation_1959, mcgibbon_identification_2017}. They are often used to decrease the dimensionality of the dynamics during the construction of Markov state models (MSM) \cite{schwantes_improvements_2013, perez_identification_2013}. Incidentally, one embarrassingly parallel strategy to exascale simulations consists of running  a very large number of short trajectories independently, which are later combined using MSMs in order to obtain a long time behavior \cite{jung_artificial_2019,wan_adaptive_2020}. 

The minimal lag time when a MSM becomes approximately Markovian, which can be estimated by the convergence of implied timescales or by Chapman-Kolmogorov criterion, is a good indicator of the accuracy of the constructed model. State of the art approaches have lag times in the range of tens of nanoseconds \cite{schwantes_improvements_2013, perez_identification_2013, hernandez_variational_2018, mardt_vampnets_2020}. Shorter lag times mean more accurate putative EVs and MSMs, as well as shorter trajectories and higher efficiency for the simple strategy of exascale simulations. Here we present an approach, which determines EVs for protein folding trajectories, which pass a stringent EV validation test at much shorter lag time of trajectory sampling interval of $0.2$ ns. 

A major difficulty of parametric approaches is that they require a functional form with many parameters to approximate RCs, e.g., linear combinations of molecular descriptors/features \cite{schwantes_improvements_2013, perez_identification_2013} or deep neural networks \cite{hernandez_variational_2018, mardt_vampnets_2020}. While, e.g., it was argued that ”the expressive power of neural networks provides a natural solution to the choice-of-basis problem” \cite{hernandez_variational_2018}, finding the optimal architecture of a neural network and input variables are difficult tasks. The suggested approach is non-parametric and can approximate any RC with high accuracy without system specific information. Instead of optimizing the parameters of the approximating function, the approach directly optimizes RC time-series. Such approaches, which use no system specific information and operate in generic, system agnostic terms, such as EVs and eigenvalues, RCs, committors  \cite{krivov_protein_2018}, optimality criteria, free energy landscapes, we propose to call blind, in analogy with the blind source separation approaches.

The blind approaches should especially be useful in the following cases: i) the initial analysis of the systems dynamics, when the knowledge of the system is very limited; ii) analyses, where one does not want to introduce any bias, e.g., due to the employed function approximation, or one does not have a satisfactory function approximation; iii) they can be used aposteriory, to check if possible bias in the analysis has altered the results.

The initial framework of non-parametric RC optimization was described in Ref.~\citenum{banushkina_nonparametric_2015}. Ref.~\citenum{krivov_protein_2018} introduced adaptive version of the approach for the committor RC to treat realistic systems with relatively limited sampling, e.g., state-of-the-art atomistic protein folding trajectories \cite{shaw_atomic-level_2010, lindorff-larsen_how_2011, piana_protein_2012}. To avoid overfitting in such a system, the approach performs RC optimization in an adaptive manner by focusing on less optimized spatiotemproal regions of RC. The latter are identified by using the committor optimality criterion \cite{krivov_reaction_2013}. The current paper makes the following contributions to the nonparametric framework. First, we suggest a rigorous and sensitive validation/optimality criterion for EVs. Second, as we discuss below, the optimization of EVs is inherently unstable \cite{banushkina_nonparametric_2015}. It is not a drawback of the non-parametric approach \textit{per se}, but is rather due to the unsupervised nature of the problem itself. One seeks EVs with the smallest eigenvalues, which describe the slowest relaxation dynamics, however, not all of such EVs are of interest. Here, we describe a few heuristics to suppress the instability. Third, we describe an adaptive approach, which avoids overfitting for realistically sampled systems. Fourth, we illustrate the power of the approach by determining accurate EVs for realistic protein folding trajectories: HP35 double mutant \cite{piana_protein_2012} and FIP35 \cite{shaw_atomic-level_2010}.  

The paper is as follows. The Methods section starts by reviewing the conventional, parametric approach of EVs approximation. Then, the non-parametric framework of RC optimization is introduced. An iterative, non-parametric approach of EV optimization is described. A stringent EV validation/optimality criterion is suggested. A protein folding trajectory is used to illustrate that iterative EV optimization has an inherently instability. It can converge to
EVs with smaller eigenvalues, but of no interest, which we denote as spurious EVs. An approach with heuristics to suppress the instability is described. Application of the EV criterion shows that
during EV optimization some regions of the putative EV are underfitted (suboptimal), while other are overfitted. The criterion is adopted to perform optimization in an adaptive, more uniform way. We conclude by discussing the obtained protein folding free energy landscapes.

\section{Method}

\subsection{Variational optimization of eigenvectors.}
Assume that system dynamics is described/approximated by a Markov chain with transition probability matrix $P(i|j,\Delta t)$ for transition from state $j$ to state $i$ after time interval $\Delta t$. Note that this assumption is used only for the derivation of equations. One does not need to know the actual Markov chain, meaning that this assumption does not restrict the applicability of the algorithm.

Given a very long equilibrium trajectory ${\bf X}(k\Delta t_0)$, where $\Delta t_0$ is the trajectory sampling interval, and using a \textit{very} fine-grained clustering of the configuration space of the system, one can, \textit{in principle}, estimate the transition matrix $P(i|j,\Delta t)=n(i|j,\Delta t)/n(j,\Delta t)$, where time interval (lag time) $\Delta t$ equals $\Delta t_0$ or its multiple, $n(i|j,\Delta t)$ is the number of transitions from cluster $j$ to cluster $i$ after time interval $\Delta t$, observed in the trajectory and $n(j,\Delta t)=\sum_i n(i|j,\Delta t)$ is the total number of transitions out of cluster $i$, which is proportional to the equilibrium probability. Knowing $P(i|j,\Delta t)$ one can estimate the left eigenvectors 
\begin{equation}
	\sum_i u'_\gamma(i) P(i|j,\Delta t)=e^{-\mu_\gamma\Delta t} u'_\gamma(j),
	\label{eq:ev1}
\end{equation}
where index $\gamma$ numbers eigenvectors, $u'_\gamma(i)$ is $\gamma$-th eigenvector as a function of cluster node ($i$), $\mu_\gamma$ is the corrsponding $\gamma$-th eigenvalue. For equilibrium dynamics with the detailed balance, $n(i|j,\Delta t)=n(j|i,\Delta t)$, which we assume here,
the smallest eigenvalue, $\mu_0=0$, and the corresponding eigenvector is constant $u'_0(j)=1$, all other eigenvalues are real and positive $\mu_{\gamma>0}>0$. 

To simplify the description of system's dynamics one can project its high-dimensional trajectory on a few EVs with lowest eigenvalues. These EVs describe slowest relaxation modes of the dynamics, and a free energy landscape as a function of these EVs can provide a simplified model of the relaxation dynamics. To project trajectory on EV $u_\gamma$ one computes EV time-series as $u_\gamma(k \Delta t_0)= u'_\gamma(i(k\Delta t_0))$; where primed variable, $u'(i)$, denotes EV as a function of cluster index $i$, $u(k\Delta t_0)$ denotes EV as a function of trajectory (trajectory snapshot number or trajectory time), while  $i(k\Delta t_0)$ denotes cluster index as a function of trajectory.

In practice, very long trajectories are rarely available, which makes this approach with accurate very fine-grained clustering non viable. Number of clusters grows exponentially with the dimensionality of the configuration space, which also limits the approach to  low-dimensional configuration space. The proposed approach determines rather accurate approximations to the time-series of a few lowest eigenvectors, $u_\gamma(k \Delta t_0)$, without performing clustering at all.

Variational approaches are a promising alternative to the clustering approach \cite{perez_identification_2013, schwantes_improvements_2013, banushkina_nonparametric_2015, banushkina_optimal_2016}. A functional form (FF) with many parameters $R(\bm{X},\alpha_i)$ (usually a weighted sum) is suggested as an approximation to EVs. One numerically optimizes the parameters by e.g., maximizing the auto-correlation function \cite{schwantes_improvements_2013,  perez_identification_2013} or minimizing the total squared displacement \cite{banushkina_nonparametric_2015}.

Namely, given a long equilibrium multidimensional trajectory $\bm{X}(k\Delta t_0)$, one computes the reaction coordinate time-series $r(k\Delta t_0)=R(\bm{X}(k\Delta t_0), \alpha_i)$. Here and below $r$ denotes any reaction coordinate, while $u$ is reserved for putative EVs. The functional form $R$ approximates the first left EV, if it provides the minimum to the total squared displacement $\Delta r^2(\Delta t)=\sum_{k=1}^{N-\Delta t/\Delta t_0}  [r(k\Delta t_0+\Delta t)-r(k\Delta t_0)]^2$, under the constraint $\sum_{k=1}^{N}  r(k\Delta t_0)^2=1$. Note that, due to the constraint, the minimization of $\Delta r^2(\Delta t)$ is equivalent to the maximization of the auto-correlation function $C(r,\Delta t)=\sum_{k=1}^{N-\Delta t/\Delta t_0} r(k\Delta t_0+\Delta t)r(k\Delta t_0)$; we neglect here small difference between constrains $\sum_{k=1}^{N} r^2(k\Delta t_0)=1$ and $\sum_{k=1}^{N-\Delta t/\Delta t_0}r^2(k\Delta t_0+\Delta t)+r^2(k\Delta t_0)=2$. The functional form $R$ approximates the $\gamma$-th left EV if it provides the minimum to the $\Delta r^2(\Delta t)$ under constraint $\sum_k r(k\Delta t_0)^2=1$ and is orthogonal to the previous $\gamma-1$ EVs $\sum_k r(k\Delta t_0)u_j(k\Delta t_0)=0$, $j=1,...,\gamma-1$.

It is straightforward to prove this principle. Consider Markov chain, describing the dynamics. Let indexes $i$ and $j$ denote the states of the chain and $r'(i)$ is an RC as a function of state $i$. Consider trajectory, i.e., a sequence of states $i(k\Delta t_0)$ of length $N$, which define the RC time-series as $r(k\Delta t_0)=r'(i(k\Delta t_0))$. The total squared displacement equals  $\Delta r^2 (\Delta t)=N\sum_{ij} [r'(i)-r'(j)]^2 P(i|j,\Delta t)P(j)$, while the constraint is $N\sum_j r'^2(j) P(j)=1$, where $P(j)$ denotes equilibrium probability. Using $2\lambda$ as the Lagrange multiplier, differentiating with respect to $r'(j)$ and assuming the detailed balance one obtains Eq. \ref{eq:ev1} with $\lambda=e^{-\mu \Delta t}$.

Consider EV time-series approximation by a linear combination of basis functions $r(k\Delta t_0)=\sum_j \alpha_j f_j(k\Delta t_0)$. Using $\lambda$ as the Lagrange multiplier the optimal values of parameters, $\alpha^\star_j$, that provide minimum to $\Delta r^2(\Delta t)$ under constraint $\sum_k r^2(k\Delta t_0)=1$ can be found as a solution of the generalized eigenvalue problem
\begin{subequations}
	\label{v:a}
	\begin{align}
		&\sum_jA_{ij}(\Delta t)\alpha_j^\star=\lambda \sum_j B_{ij}\alpha_j^\star \\
		&A_{ij}(\Delta t)=\sum_{k=1}^{N-\Delta t/\Delta t_0} \Delta f_i(k\Delta t_0)\Delta f_j(k\Delta t_0) \\
		&B_{ij}=\sum_{k=1}^{N} f_i(k\Delta t_0)f_j(k\Delta t_0),
	\end{align}
\end{subequations}
where $\Delta f_i(t)=f_i(t+\Delta t)-f_i(t)$ denotes the forward time difference. 
The solutions of the eigenvalue problem are found numerically by standard linear algebra methods. Since both matrices are symmetric, the eigenvalues are real. Assume that eigenvalues are sorted as $\lambda_0=0<\lambda_1<\lambda_2...$.
Then, the $\gamma$-th solution of Eq. \ref{v:a}a, denoted as  $\alpha^{\star\gamma}_j$, corresponds to putative RC time-series $r(k\Delta t_0)=\sum_j \alpha_j^{\star\gamma} f_j(k\Delta t_0)$, which approximates $\gamma$-th EV time-series $u_\gamma(k\Delta t_0)$

\subsection{Estimation of eigenvalues and implied timescales}
\label{impts}
The minimal value of the $\Delta r^2(\Delta t)$ functional, attained when $r$ approximates EV $u$, equals $\Delta u^2(\Delta t)=2(1-e^{-\mu \Delta t})$, which, for small $\Delta t$, gives $\Delta u^2(\Delta t)\approx 2\mu \Delta t$; it is assumed here that EV is normalized as $\sum_k u^2(k\Delta t_0) = 1$. Correspondingly, the maximum value of the auto-correlation term equals $C(u,\Delta t)=e^{-\mu\Delta t}$. They can be used to estimate the eigenvalues $\mu$, or the so called implied timescales $\hat{\tau}=1/\mu$ as
\begin{subequations}
	\label{eq:tau}
	\begin{align}
		\mu&=-\ln[1-\Delta u^2(\Delta t)/2]/\Delta t \label{eq:tau1}\\
		\mu&=-\ln[C(u,\Delta t)]/\Delta t \label{eq:tau2}
	\end{align}
\end{subequations}
as functions of lag time $\Delta t$. Large lag times mask suboptimality of the putative EV and lead to a more accurate estimates of $\mu$ and $\hat{\tau}$. However at very large lag times it becomes difficult to accurately estimate an exponentially decreasing value of $C(u,\Delta t)=e^{-\mu\Delta t}$, since its statistical accuracy is limited by the number of transitions between regions where $u$ is positive and negative, i.e., different free energy minima. An accurate and robust estimate should have statistical errors much smaller than the estimated value. A characteristic lag time $\Delta t^\star$, where the two are comparable could be roughly estimated as  $(\mu T)^{-1/2}= e^{-\mu\Delta t^\star}$, where $T$ is the total duration of the trajectory. The lag time chosen to accurately estimate the eigenvalues and the implied timescales, which we denote as $\Delta t_\infty$, should be chosen much smaller than $\Delta t^\star$.
An EV optimization is considered to be converged when eigenvalue estimated with lag time of interest $\Delta t$ is close to the accurate eigenvalue, i.e., $\mu(\Delta t)\approx \mu(\Delta t_\infty)$.

In application to the HP35 protein, considered here, $\Delta t^\star\sim 10^4 \Delta t_0$ and we took $\Delta t_\infty$ ten times smaller,  $\Delta t_\infty=1024\Delta t_0=204.8$ ns as $\Delta t_0=0.2$ ns. The described approach determines EVs with eigenvalues (and implied timescales) accurate at the lag time of trajectory sampling interval of $0.2$ ns, i.e.,  $\mu(\Delta t_0)\approx \mu(\Delta t_\infty)$.

\subsection{Non-parametric optimization of eigenvectors}
A major weakness of parametric approaches that approximate RCs by using a functional form (FF) with many parameters, e.g., a linear combination of collective variables \cite{} or a neural network, is that it is difficult to suggest a good FF approximating EVs. The difficulty becomes apparent if one remembers that such a FF should be able to accurately project a few million snapshots of a very high-dimensional trajectory. In particular, it implies an extensive knowledge of the system, and that such a FF is likely to be system specific. 

Recently we have suggested a non-parametric approach for the determination of the committor function, which bypasses the difficult problem of finding an appropriate FF \cite{banushkina_nonparametric_2015, krivov_protein_2018}. The power of the approach was demonstrated by applying it to the equilibrium folding trajectory of the HP35 double mutant. The determined RC closely approximates the committor as was validated by the optimality criterion - $Z_{C,1}$ (defined below) is constant up to the expected statistical noise  \cite{krivov_reaction_2013}. The approach performs optimization of the RC in a uniform manner by focusing optimization on the time scales and the regions of the putative RC which are most suboptimal. 

The general idea of iterative non-parametric RC optimization is as follows \cite{banushkina_nonparametric_2015, krivov_protein_2018}. We start with a seed RC time-series $r(k\Delta t_0)$. During each iteration we consider a variation of RC as $r(k\Delta t_0)+\delta r(k\Delta t_0)$, where $\delta r(k\Delta t_0)$ can be a time-series of any function of configuration space, collective variables and, hence, the RC itself. For example, one can take $\delta r(k\Delta t_0) = f (r(k\Delta t_0), y(k\Delta t_0))$, where $y(k\Delta t_0)$ is time-series of a randomly chosen collective  variable or a coordinate of the configuration space and $f(r,y)=\sum_{ij} \alpha_{ij} r^i y^j$ is a low degree polynomial. The coefficients/parameters of the variation are chosen such that $r(k\Delta t_0) + \delta  r(k\Delta t_0)$ provides the best approximation to the
target optimal RC (e.g., the committor or EVs). Specifically, they deliver optimum to the corresponding target functional. For the optimization of EVs, considered here, they can be found as solutions of Eq. \ref{v:a}. The RC time-series is updated $r(k\Delta t_0) \leftarrow r(k\Delta t_0) + \delta r(k\Delta t_0)$ and the process is repeated. Iterating the process
one repeatedly improves the putative RC time-series by incorporating information contained in different coordinates or collective variables. The process stops when, e.g., the target functional is close to its optimal value, meaning that the putative RC is a close approximation of the target RC. 

Importantly, while the result of each iteration may depend on the exact choice of the family of collective variables or the functional form of the variation, the final RC does not, since it provides the optimum to a (non-parametric) target functional when the optimization converges, which makes this approach non-parametric. It is assumed that the family of collective variables contains all the important information about the dynamics of interest. If the system obeys some symmetry (e.g., the rotational and translational symmetries for biomolecules), then the RCs should obey the same symmetry. A simple way to ensure this is to use collective variables that respect the symmetry. For example, the distances between randomly chosen pairs of atoms or $\sin$ and $\cos$ of dihedral angles can be suggested as standard sets of collective variables.

Here we extend the approach to non-parametric determination of eigenvectors. Specifically, given a multidimensional trajectory $\bm{X}(k\Delta t_0)$ and the number of the slowest eigenvectors required $n_\mathrm{ev}$, the approach determines time-series of the required eigenvectors $u_\gamma(k\Delta t_0)$ and corresponding eigenvalues $\mu_\gamma$, where $k=[1,N]$ and $N$ is the trajectory length and $1\le \gamma\le n_\mathrm{ev}$.

We start with seed EVs time-series, $u_\gamma(k\Delta t_0)$, $1\le \gamma\le n_\mathrm{ev}$, for example the distance time-series between randomly chosen pairs of atoms. Then, the EVs time-series are improved iteratively. To simultaneously update all EVs during each iteration, we consider a variation of EVs time-series as 
\begin{equation}
	r(k\Delta t_0)=\sum_{\gamma} \alpha_\gamma u_\gamma(k\Delta t_0)+f(u_\beta(k\Delta t_0), y(k\Delta t_0)),
	\label{eq:iter}
\end{equation}
here, $y(k\Delta t_0)$ is the time-series of a randomly chosen collective variable of the original multidimensional space $\bm{X}$, $\beta$ denotes index of an active EV, whose contribution to the variation is higher then linear, and $f(u,y)=\sum a_{ij} u^i y^j$ is a low degree polynomial. 

All the time-series in the variation (Eq. \ref{eq:iter}) are denoted as basis functions $f_j(k\Delta t_0)$; the variation can be written as $r(k\Delta t_0)=\sum_j \alpha_j f_j(k\Delta t_0)$, where vector $\alpha_j$ now contains both parameters $\alpha_\gamma$ and coefficients of the polynomial $a_{ij}$. The optimal values of the parameters, $\alpha_j^\star$, are chosen such that the variation provides the best approximation to an EV time-series. They are determined by numerically solving 
Eq. \ref{v:a}. The first $n_\mathrm{ev}$ solutions,  denoted as  $\alpha^{\star\gamma}_j$, are used to update the putative time-series of $\gamma$-th EV as $u_\gamma(k\Delta t_0)\leftarrow\sum_j \alpha_j^{\star\gamma} f_j(k\Delta t_0)$, and the iterative process is repeated.

The generalized eigenvalue problem, Eq. \ref{v:a}, does not have a solution if the basis functions contain the same time-series twice. Here, the time-series of the active EV, $u_\beta(k\Delta t_0)$, is included in both the first sum and the polynomial. The same is true for EV $u_0$, corresponding to $\mu_0=0$, whose time-series is a constant. To have these time-series only once we assume that the constant and $u$ terms are removed from the polynomial.

Inclusion of the linear combination of all EVs into the RC variation (Eq. \ref{eq:iter}) means that this variation can be considered as a variation $u_\gamma+\delta u$ of every EV in turn. It ensures, in particular, that every updated EV has the corresponding EV at the previous iteration as a baseline. Active EVs can be selected randomly, or one may select the least optimal EV, i.e., the one having the largest ratio  $\mu(\Delta t)/\mu(\Delta t_\infty)$. The iterative optimization is considered to be converged, when eigenvalues of all eigenvectors of interest estimated with the lag time of interest $\Delta t$ are close to the accurate values, i.e., $\mu_\gamma(\Delta t)\approx \mu_\gamma(\Delta t_\infty)$.

Thus, a minimal algorithm of non-parametric EV optimization is as follows. 
\textbf{Initialization:} Set seed EVs time-series. $u_0(k\Delta t_0)=1$. For $1\le \gamma \le n_\mathrm{ev}$ select randomly a collective variable $y$ and set $u_{\gamma}(k\Delta t_0)=y(k\Delta t_0)$. Select the lag time of interest $\Delta t$ and the lag time $\Delta t_\infty$ to test convergence. For example,  $\Delta t=\Delta t_0$ and $\Delta t_\infty=1024 \Delta t_0$.
\textbf{Iterations:}
Select active EV, $u_\beta$, as the most suboptimal one, i.e., the one with the largest ratio $\beta=\arg \max_\gamma \mu_\gamma(\Delta t)/\mu_\gamma(\Delta t_\infty)$, or just randomly. Select collective variable time-series $y(k\Delta t_0)$. Compute basis functions of Eq. \ref{eq:iter}, solve Eq. \ref{v:a} and updates the EVs time-series $u_\gamma(k\Delta t_0)$.
\textbf{Stopping:} 
Stop if the optimization has converged: $\mu_\gamma(\Delta t) < \mu_\gamma(\Delta t_\infty)$ for $1\le \gamma \le n_\mathrm{ev}$.

To explicitly illustrate the iterative character of the optimization, the algorithm can be written as  $u^{n+1}_1,...,u^{n+1}_{n_{ev}}=F(u^{n}_1,...,u^{n}_{n_{ev}},\beta^n,y^n)$, where superscript $n$ denotes values of variables at $n$-th iteration, and $F(u^{n}_1,...,u^{n}_{n_{ev}},\beta,y)$ denotes a function/procedure that takes a set of $n_{ev}$ EVs time-series $u_\gamma$, the index of active eigenvector $\beta$ and time-series of collective variable $y$, computes basis functions of Eq. \ref{eq:iter}, solves Eq. \ref{v:a} and returns a set of updated $n_{ev}$ time-series $v_\gamma$, that better approximate the EVs.

\textbf{Selecting collective variable} time-series $y(k\Delta t_0)$ means random selection from the provided set of collective variables. For example, if one takes a standard set of collective variables - the inter-atom distances, then every time a collective variable is requested, one selects a random pair of atoms $i$ and $j$, and returns the distance time-series between the atoms $r_{ij}(k\Delta t_0)$ computed from the trajectory.

\textbf{Selection of $\Delta t_\infty$.} Lag time $\Delta t_\infty$ is used to test the convergence of EV optimization as $\mu_\gamma(\Delta t) \approx \mu_\gamma(\Delta t_\infty)$. From one side, it should be chosen as long as possible, to mask the deficiencies of putative EV time-series and have a more accurate estimate of eigenvalue $\mu_\gamma$. From the other side, very long $\Delta t_\infty$ lead to large statistical uncertainties in the estimation of $\mu_\gamma(\Delta t_\infty)$, as discussed in Sect. \ref{impts}. One strategy of selection of $\Delta t_\infty$ is to, first, perform optimization with $\Delta t_\infty$ conservatively selected just a few times longer than the lag time of interest $\Delta t$, determine the statistical uncertainties as a function of lag time using bootstrapping and use that for an informed selection of $\Delta t_\infty$.

\textbf{Selection of the polynomial.} Generally, the higher is the degree of the polynomial $f(u,y)$, the faster is the optimization, though more computationally demanding. However a very high degree may lead to numerical instabilities and strong overfitting. The following strategy was found useful: use a polynomial $f(u,y)$ with a relatively small degree (3-6) for updates involving $u_\beta$ and $y$ followed by a polynomial $f(u)$ of a high degree (e.g., 10-16) for updates involving only $u_\beta$, where $u_\beta$ is the active EV.
 
\subsection{Eigenvector validation/optimality criterion}
An accurate eigenvalue or the corresponding implied timescales can serve as an indicator that the putative RC time-series closely approximates an EV.  
However, these metrics provide a rather crude, cumulative estimate of the accuracy of putative EVs. It is possible that while an eigenvalue is accurate, some parts of EV are overfitted/overoptimized, while other underfitted. To check for that, we describe a more stringent EV optimality/validation criterion $\Theta(x,\Delta t)$. 

The criterion is an extension of the $Z_{C,1}$ criterion for the committor reaction coordinate. $Z_{C,1}$ can be straightforwardly computed from time-series $r(k\Delta t_0)$:  each transition of trajectory from $x_1=r(i\Delta t)$ to $x_2=r(i\Delta t+\Delta t)$ adds $1/2 |x_1-x_2|$ to $Z_{C,1}(x,\Delta t)$ for all points $x$ between $x_1$ and $x_2$ \cite{krivov_reaction_2013, krivov_protein_2018}. Jupyter notebooks illustrating usage of $Z_{C,\alpha}$ profiles and the committor and eigenvector criteria are available at  \href{https://github.com/krivovsv/CFEPs}{https://github.com/krivovsv/CFEPs} \cite{CFEPs}.

$Z_{C,1}$ has a number of useful properties \cite{krivov_reaction_2013, krivov_protein_2018}. If reaction coordinate $q$ closely approximates the committor function, then $Z_{C,1}(q, \Delta t)\approx N_{AB}$, where $N_{AB}$ is the number of transitions between boundary states, i.e., from $A$ to $B$, or from $B$ to $A$. For a suboptimal reaction coordinate $r$, $Z_{C,1}(r, \Delta t)$ values generally decrease to the limiting value of $N_{AB}$, as $\Delta t$ increases. The larger the difference between $Z_{C,1}(r, \Delta t_1)$ and $Z_{C,1}(r, \Delta t_2)$ the less optimal the reaction coordinate around r. This property is used to find suboptimal spatio-temporal regions and focus optimization on them to make it more uniform.

The constancy of $Z_{C,1}(q,\Delta t)$ along the committor $q$ follows from the following. Consider Markov chain, describing the dynamics. Let indexes $i$ and $j$ denote the states of the chain and $x(i)$ their position on an RC. Value of $Z_{C,1}(x,\Delta t)$ can change, in a step-wise fashion, only when position $x$ goes through a particular state $j$, i.e., $x$ goes from $x(j)-0$ to $x(j)+0$ and equals \cite{krivov_reaction_2013} 
\begin{equation}
	\Delta Z_{C,1}(x(j),\Delta t)=\sum_i [x(i)-x(j)]n(i|j,\Delta t).
\end{equation}
It is zero for the committor function (if $j$ is not a boundary state) since committor is defined by the following equation
\begin{subequations}
	\begin{align}
		&\sum_i [q(i)-q(j)] P(i|j,\Delta t)=0 \quad \mathrm{for} \, j \ne A, B \label{eq:qeq}\\
		&q(A)=0,\quad q(B)=1 \label{eq:qbc}
	\end{align}
\end{subequations}
and $n(i|j,\Delta t)=P(i|j,\Delta t)P(j)$. Eq.  \ref{eq:ev1} is different from \ref{eq:qeq} which means that $Z_{C,1}$ along an eigenvector is not constant. However Eq. \ref{eq:ev1} can be rewritten as 
\begin{equation}
	\sum_i[u'(i)-u'(j)]n(i|j,\Delta t)=(1-e^{-\mu\Delta t})[0-u'(j)]n(j),
	\label{eq:ev2}
\end{equation}
and interpreted in the following way. On the left hand side we have change in $Z_{C,1}$ around $u'(j)$ computed in the standard way. It is proportional to the change of $Z_{C,1}$ computed for a virtual trajectory consisting of collection of transitions $0$ to $u'(j)$ and back to $0$ made $n(j)$ times for every $j$. We denote the second profile as $Z^0_{C,1}$. Since both profiles are $0$ at large negative $x$ and have proportional changes, they are proportional themselves $Z_{C,1}(x,\Delta t)=(1-e^{-\mu\Delta t})Z^0_{C,1}(x,\Delta t)$.
Note that $Z^0_{C,1}(x,\Delta t=m\Delta t_0)=Z^0_{C,1}(x,\Delta t_0)/m$.
Consider the following variable 
\begin{equation}
	\Theta(x,\Delta t)=-\ln\frac{Z_{C,1}(x,\Delta t)}{(1-e^{-\mu\Delta t})Z^0_{C,1}(x,\Delta t)}.
\end{equation}

\textbf{Validation:} If putative time-series $u(i \Delta t_0)$ and $\mu$ closely approximates an EV and the corresponding eigenvalue, then $\Theta(x,\Delta t)\approx 0$ for all $\Delta t$ and all $x$ along $u$.
An accurate estimate of $\mu$ is obtained from the EV time-series using Eq. \ref{eq:tau} at large lag times.

$Z_{C,1}(x,\Delta t)$ can be interpreted as a local density of the total squared displacement $\Delta r^2(\Delta t)/2$, since $\int Z_{C,1}(x,\Delta t)dx=\Delta r^2(\Delta t)/2$ \cite{krivov_reaction_2013}. Analogously, $Z^0_{C,1}(x)$ can be considered as a local density of $\sum_k r^2(k\Delta t_0)$. The constraint optimization problem is equivalent to finding minimum of an integral of $Z_{C,1}(x,\Delta t)$ under constraint that an integral of $Z^0_{C,1}(x)$ is $1$. When a putative coordinate closely approximates an eigenvector, the local densities are proportional. 

\textbf{Optimality:} For a suboptimal coordinate 
$\Theta(x,\Delta t)<0$, because $Z_{C,1}(x, \Delta t)$ is larger than that for the optimal coordinate. The bigger the difference between $\Theta(x,\Delta t_1)$  and $\Theta(x,\Delta t_2)$ for $t_1>t_2$ the less optimal is $u$ around $x$.  $\Theta(x,\Delta t)\rightarrow 0$ as $\Delta t$ increases.

\subsection{Inherent instability of iterative EV optimization}
We illustrate the minimal algorithm of non-parametric optimization by determining the first eigenvector $u_1$ for a long equilibrium trajectory of double mutant of HP35 protein consisting of 1509392 snapshots at 380 K and the sampling interval of $\Delta t_0=0.2$ ns \cite{piana_protein_2012}. We used $\Delta t=\Delta t_0$, $\Delta t_\infty=1024\Delta t_0$, and polynomials $f(u_1,y)$ of degree 4 and $f(u_1)$ of degree 12. Inter-atom distance were used as a set of collective variables.

\begin{figure}[htbp]
	\centering
	\includegraphics[width=.5\linewidth]{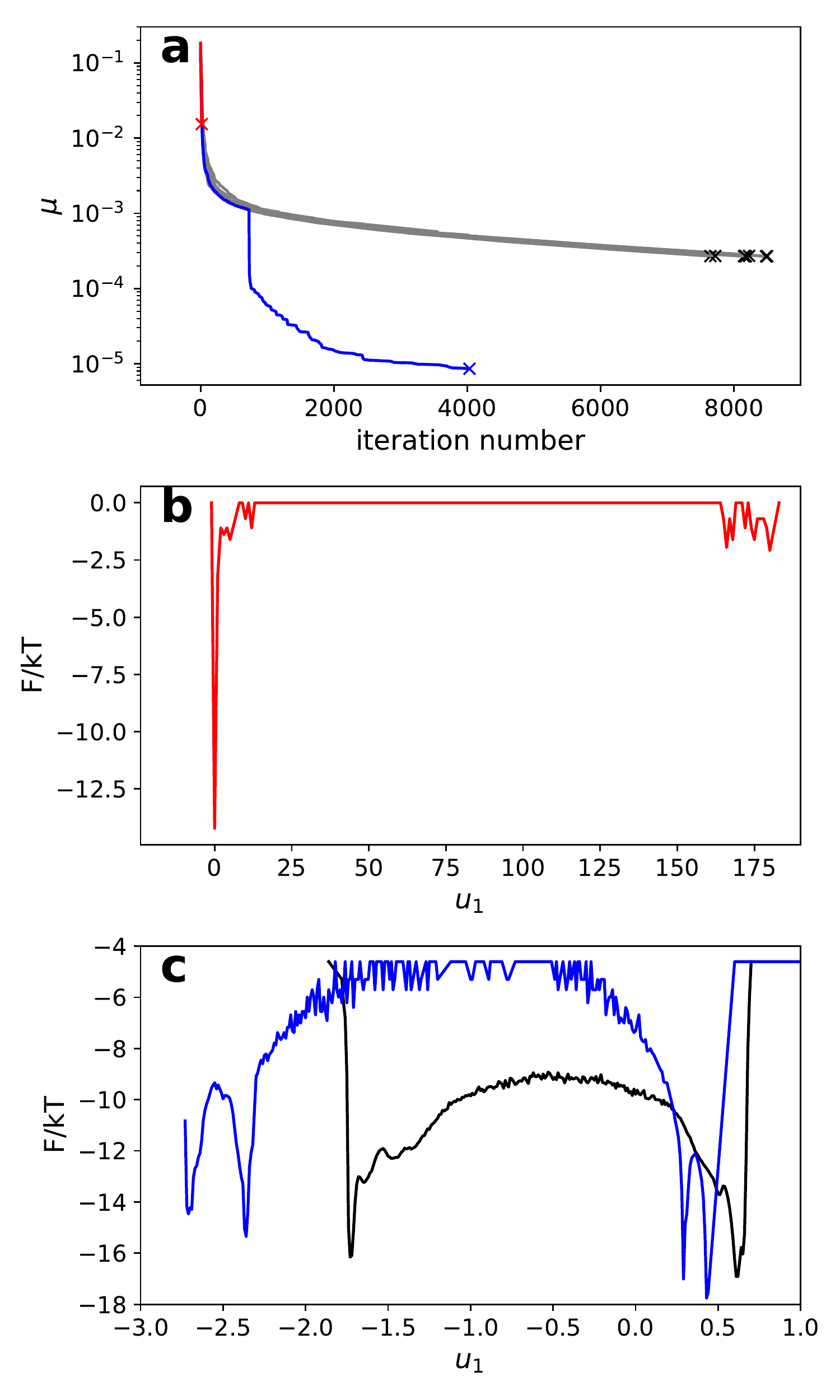}
	\caption{Application of the minimal algorithm of non-parametric EV optimization to the determination of $u_1$ of HP35. Ten representative optimization runs started with different seed numbers. {\bf a}) Estimate of the eigenvalue $\mu(\Delta t_0$) in units of $\Delta t_0^{-1}$ as a function of iteration number; cross at the end of line indicates where optimization has converged, i.e., $\mu_\gamma(\Delta t_0)\approx \mu_\gamma(\Delta t_\infty)$. Most of the lines (gray) fall on the same curve, indicating the robustness and reproducibility of the non-parametric approach. For them the eigenvalues steadily decrease toward the target value. However, two optimization runs converged to different, spurious eigenvectors (see text). Run, colored run, converged to a EV, which FEP, $F(u_1)$, is shown on {\bf b}).
	Optimization run, colored blue, initially followed the common trajectory and deviated abruptly around 720-th iteration and converged to an EV with lower eigenvalue. Free energy profiles of putative EVs just before the iteration and straight after are shown on {\bf c}) by black and blue colors, respectively.}
	\label{fig:spur}
\end{figure}

Fig. \ref{fig:spur}a shows $\mu(\Delta t_0)$ as a function of iteration number for ten representative optimization runs started with different random seed numbers. For most of the runs $\mu(\Delta t_0)$ steadily converges to the same eigenvalue of $\mu(\Delta t_0)\sim 2.68 \cdot 10^{-4}$ in units of $\Delta t_0^{-1}$. It indicates robustness and reproducibility of the non-parametric optimization. The putative time-series, after a few thousands iterations, can provide a rather good approximation to an EV with the corresponding eigenvalue within a small factor from the exact value.

However, two of the runs, showed by red and blue colors, converged to different EVs with different eigenvalues.  Optimization run, showed by red color on Fig. \ref{fig:spur}a is rather short. It started as other runs but quickly converged to a spurious EV. The EV has a peculiar free energy profile (FEP), $F(u_1)$, shown on Fig. \ref{fig:spur}b; the FEP is estimated from a histogram. The EV has a rather large amplitude $A(u_1)=\max(u_1)-\min(u_1)\approx 175$ and describes a transition to a low populated,  shallow minimum. Inspection of the EV time-series shows that it has made only one transition from the main minimum around $u_1\sim 0$ to the shallow minimum around $u_1\sim 175$ and back. Optimization run, showed by blue color, initially followed the gray lines, however around 720-th iteration in deviated abruptly, which can be seen by the abrupt change of the eigenvalue on Fig. \ref{fig:spur}a. FEP of the putative EV just before the iteration is shown by black line on Fig. \ref{fig:spur}c, and is very close to the FEPs of EVs the runs, colored gray, converged to. The blue line on Fig. \ref{fig:spur}c shows the FEP of the putative EV just after the abrupt transition, which has a much higher barrier and more structure. Correspondingly, the EV has a much smaller eigenvalue of $\mu(\Delta t_0)\sim 8.58\cdot 10^{-6}$ in units of $\Delta t_0^{-1}$. However, the FEP does not describe the folding dynamics. The two minima of the FEP, $u_1<-1$ and $u_1>-1$, have identical FEPs when projected on the root-mean-square-deviation  from the native structure RC. Closer inspection shows that the main barrier describes a rotation of a dihedral angle corresponding to a transition between two permutational isomers of GLN 67 residue. Collective variable $y=r_{ij}$ that contributed to this abrupt deviation is the distance between atoms  209 and 491, which correspond to  OE1 and HA4 in GLN 67. The permutational isomers correspond to exchange of hydrogen atoms HA4 and HE41. Thus, while this EV has a smaller eigenvalue, it has no connection to folding and is of very limited interest.

To summarize, the two deviated runs illustrate that the problem of determining the slowest EVs has the following inherent "instability" \cite{banushkina_nonparametric_2015}. The algorithm seeks EVs with the smallest eigenvalues, which describe the slowest dynamics. However, some of such EVs, which we denote as spurious EVs, are not of interest. For example, in protein folding, such an EV could describe a much slower torsion angle isomerization process  \cite{banushkina_nonparametric_2015, mcgibbon_identification_2017}. The spurious EV shown by blue color on Fig. \ref{fig:spur}, describes a permutational isomerization, that happened 7 times in the course of the entire trajectory that contains about 140 folding-unfolding events. Another, more frequent possibility, is due to limited sampling. There are many parts of the configuration space that were visited very few times or even just once (the shallow basin on Fig. \ref{fig:spur}b), and EVs describing those transitions have small eigenvalues. Thus, starting with an EV of interest, the algorithm may eventually converge to a spurious EV, with smaller eigenvalue, but of no interest. In general terms, this peculiarity of EV optimization is due to its unsupervised nature: we seek any EV with smallest eigenvalue. Optimization of the committor function, which is a variant of supervised learning, as the function interpolates between two given boundary states of interest, is free of such a problem \cite{banushkina_nonparametric_2015, krivov_protein_2018}.

For systems, where the likelihood of switching to spurious EVs is not large, the simplest is to just discard the optimization runs that have converged to spurious EVs, and keep those where EVs of interest are found. Such systems can be analyzed with the minimal algorithm described above. For other systems, where the likelihood of switching to spurious EVs is large, one needs a more systematic approach of suppressing the instability. We describe a few heuristics to suppress the instability, which were sufficient to determine the lowest EVs of realistic protein folding trajectories.

As illustrated on Fig. \ref{fig:spur}, a shift to a spurious EV happens usually in an abrupt manner and results in significant changes in the EV time-series. Thus, allowing only gradual changes of the putative EV time-series. should help suppress the instability.
A main idea is to keep a fraction of trajectory points, selected with probability $p_\mathrm{fix}$ ($0.5$ here), fixed during each iteration. It penalizes large changes in the EV time-series, since during optimization the distance between consecutive points is minimized. Allowing, an overall shift and change of scale, it means that fixed points are transformed according to Eq. \ref{eq:iter}, with contributions from the polynomial set to zero, i.e., all eigenvectors contribute linearly. Increasing $p_\mathrm{fix}$ enforces a more gradual change of eigenvectors during optimization.

The eigenvalue of an EV, estimated at large lag time $\Delta t_\infty$, changes rather little after an initial settling phase. Hence, a relatively large change ($5$ \% here), is an indication that an EV has changed significantly.  Iterations with such changes are not accepted.

Collective variables that promote transitions to spurious EVs,
(e.g., like that on Fig. \ref{fig:spur}c) can be filtered out. A simple collective variable $y$ that depends on a few coordinates only, e.g., the inter-atom distance, is first transformed to the first EV as its function $y\rightarrow u_1(y)$. If the corresponding eigenvalue is very small it means that $y$ does not describe a collective process, such as protein folding, and is likely to describe a spurious EV. Such a variable is discarded.

In the infrequent cases, when, in spite of the heuristics employed, the algorithm switches to a spurious EV, the optimization is restarted. Such events are detected by the following heuristics: one monitors the amplitude of an EV $A(u)$. When the amplitude reaches a relatively large value, it indicates of a spurious EV analogous to that on Fig. \ref{fig:spur}b; e.g., compare the amplitudes of EVs on Fig. \ref{fig:spur}b and Fig. \ref{fig:spur}c. 

Note that, usually, the likelihood of switching to a spurious EV from the very start is rather small. Thus with a large likelihood a randomly selected collective variable will naturally lead to the slowest EVs describing a collective process, like protein folding, i.e., the EVs of interest. There is no need to specifically select an EV of interest which keeps the analysis unbiased and blind. 

The optimization algorithm with heuristics to suppressed instability is as follows.  \textbf{Initialization:} Set seed EVs time-series. $u_0(k\Delta t_0)=1$. For $1\le \gamma \le n_\mathrm{ev}$ select randomly a collective variable $y$ and set $u_{\gamma}(k\Delta t_0)=y(k\Delta t_0)$. Set the starting lag time $\Delta t>\Delta t_0$ and the lag time $\Delta t_\infty$ to test convergence. For example,  $\Delta t=256\Delta t_0$ and $\Delta t_\infty=1024 \Delta t_0$. Set the $p_\mathrm{fix}$ probability, e.g., $p_\mathrm{fix}=0.5$.
\textbf{Iterations:}
Select the set of fixed points with probability $p_\mathrm{fix}$.
Select active EV, $u_\beta$, as the most suboptimal one, i.e., the one with the largest ratio $\beta=\arg \max_\gamma \mu_\gamma(\Delta t)/\mu_\gamma(\Delta t_\infty)$, or just randomly. Select randomly collective variable $y$. Compute basis functions of Eq. \ref{eq:iter}. Set polynomial basis functions to $0$ for fixed points/frames. Solve Eq. \ref{v:a} and compute updates for the EVs.
If the update passes the safety checks for the suppression of the instability, update the EVs. If optimization has diverged: an EV amplitude $A(u)$ has crossed the threshold (30 here), restart the optimization.
\textbf{Stopping:} Optimization with current lag time stops when the eigenvalue estimate is close to the accurate value $\mu_\gamma(\Delta t) < \mu_\gamma(\Delta t_\infty)$. If $\Delta t>\Delta t_0$, then $\Delta t$ is halved and optimization with smaller $\Delta t$ is continued. If $\Delta t=\Delta t_0$ optimization stops. 

\textbf{Selection of collective variables.} To filter out collective variables that promote transitions to spurious EVs one proceeds as follows. Select a random pair of atoms $i$ and $j$, and compute the distance time-series between the atoms $y(k\Delta t_0)=r_{ij}(k\Delta t_0)$ from the trajectory. Compute $u_1(y)$ and the corresponding eigenvalue, using the Eq. \ref{v:a} with basis functions $f_j(k\Delta t_0)=y^j(k\Delta t_0)$ for $0\le j\le 12$. If eigenvalue is smaller than a threshold (e.g., $\mu(\Delta t)<10^{-4}$ here), reject $y$ and repeat the process with another pair of atoms.

\textbf{Selection of  $p_\mathrm{fix}$.} The larger is $p_\mathrm{fix}$ the more robust, but slower is optimization. One is advised to start with $p_\mathrm{fix}=0.5$ and adjust according to the performance of the algorithms.

\begin{figure}[htbp]
	\centering
	\includegraphics[width=.5\linewidth]{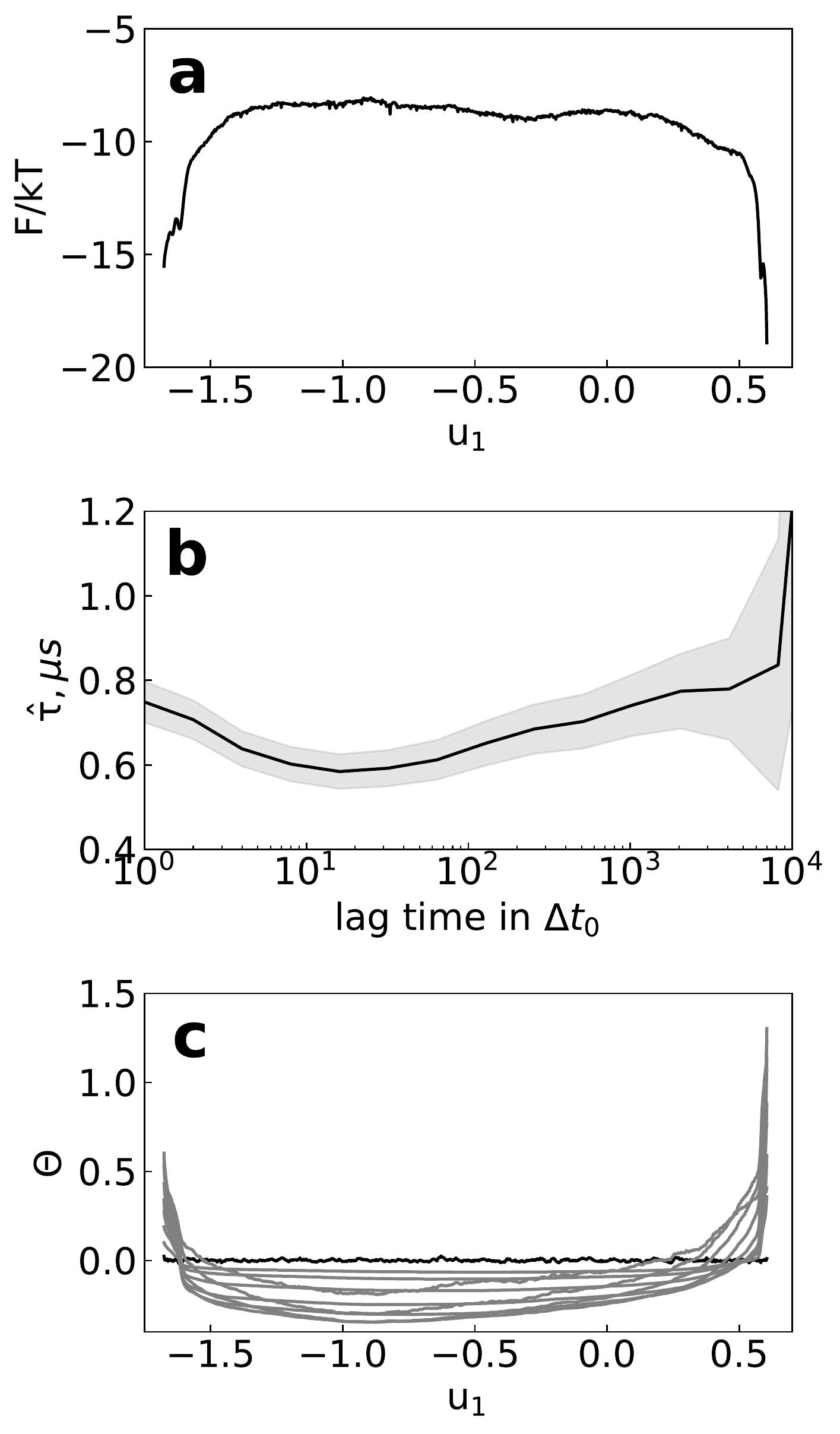}
	\caption{Non-parametric optimization of first eigenvector $u_1$ of HP35 with heuristics to suppress instability. {\bf a}) Free energy as a function of $u_1$. {\bf b}) Implied timescale $\hat{\tau}$ as a function of lag time $\Delta t$. Uncertainties (shaded areas) were computed with bootstrap. Uncertainties rapidly increase as the lag time approaches $\Delta t^\star$. {\bf c}) Optimality criterion for an eigenvector $\Theta(u_1,\Delta t)$ for different lag times $\Delta t/\Delta t_0=[1,2,4, ..., 2^{10}]$. $\Theta(u_1,\Delta t_0)$ is shown by solid black line.}
	\label{fig:non}
\end{figure}

Fig. \ref{fig:non} shows the application of the algorithm with the suppressed instability to the determination of $u_1$ of HP35. Fig. \ref{fig:non}a shows FEP as a function of the first EV $F(u_1)$. It has a simple shape of one free energy (FE) barrier and two minima. 

Fig. \ref{fig:non}b shows that an accurate estimate of implied timescales is possible with lag time of the trajectory sampling interval of $\Delta t_0=0.2$ ns. The figure confirms the choice of $\Delta t_\infty=1024 \Delta t_0$. It is sufficiently long, so that the estimates of the eigenvalue or implied timescale at this lag time agrees with those at longer lag times. At the same time it is sufficiently short so that the estimates have small uncertainty. 

The uncertainties of the estimate of implied timescales rapidly increase as $\Delta t$ approaches $\Delta t^\star$.
As explained in Sect. \ref{impts} it is difficult to accurately estimate the exponentially decreasing auto-correlation function $C(r,\Delta t)\approx e^{-\mu \Delta t}$. $C(r,\Delta t)$ decreases as lag time increases because a larger fraction of points transit to the other basin with the opposite sign of EV. The statistical error of the estimate of $C(r,\Delta t)$ is determined by the total number of transitions between the basins. And when, with increasing lag time, the estimate of $C(r,\Delta t)$ becomes close to its statistical error, the uncertainty of $\mu_1$ estimate rapidly increases. 

Fig. \ref{fig:non}c shows the EV optimality/validation criterion $\Theta(x,\Delta t)$. In particular, it shows that $\Theta(x,\Delta t_0)>\Theta(x,\Delta t)$ around the barrier and $\Theta(x,\Delta t_0)<\Theta(x,\Delta t)$ around minima for large $\Delta t>\Delta t_0$. It means that the putative $u_1$ time-series does not approximate the EV uniformly. It overfits the EV around the barrier region and underfits around the minima.

To conclude, while the accurate eigenvalue suggest that the putative time-series closely approximates $u_1$, the more stringent EV optimality/validation criterion shows that the time-series overfits $u_1$ in some parts and underfits in other.

\subsection{Adaptive optimization}
Our aim is to determine such an EV time-series $u(i\Delta t_0)$ that it passes the validation test, i.e., $\Theta(u,\Delta t)\approx 0$ up to statistical uncertainty. A way to do this is to perform optimization more uniformly, so that all regions of the putative EVs become underfitted to the same degree and stop optimization just before overfitting. Such an adaptive optimization is performed by focusing on less optimized parts of putative EVs. Before every iteration one scans $\Theta(x,\Delta t)$ profiles to find most suboptimal/underfitted regions. Position dependent $p_\mathrm{fix}(x)$ is introduced in such a way as to be smaller for more underfitted regions. Smaller $p_\mathrm{fix}(x)$ means less constraints and thus faster optimization. The obtained results are robust with respect to specific form of $p_\mathrm{fix}(x)$ employed.  More details are given in the Appendix. 

The generic adaptive non-parametric EV optimization algorithm is as follows.
\textbf{Initialization:}
Set seed EVs time-series. $u_0(k\Delta t_0)=1$. For $1\le \gamma \le n_\mathrm{ev}$ set $u_{\gamma}(k\Delta t_0)=y(k\Delta t_0)$, where y is a randomly selected collective variable, e.g., from the standard set.
Set the initial lag time to a large value, e.g., $\Delta t=256 \Delta t_0$. Set $\Delta t_\infty$, for example, $\Delta t_\infty=1024 \Delta t_0$
\textbf{Iterations:}
\begin{enumerate}
\item Select active EV, $u_\beta$, as the most suboptimal one, i.e., the one with the largest ratio $\beta=\arg \max_\gamma \mu_\gamma(\Delta t)/\mu_\gamma(\Delta t_\infty)$, or just randomly.
\item Scan $\Theta(x,\Delta t)$ profiles for the active EV to find most suboptimal/underfitted regions and compute the position dependent $p_\mathrm{fix}(x)$. Determine fixed points/frames: for frame $k$ take the position of the frame along the active EV, $x=u_\beta(k\Delta t_0)$, and choose the frame to be fixed with probability $p_\mathrm{fix}(x)$.
\item Select randomly collective variable $y$. Compute basis functions of Eq. \ref{eq:iter}. Set polynomial basis functions to $0$ for fixed points/frames. Solve Eq. \ref{v:a} and compute updates for the EVs.
\item Perform safety checks. If optimization has diverged, an eigenvector amplitude $A(u)=\max(u)-\min(u)$ has crossed the threshold (30 here), restart the optimization by going to \textbf{Initialization}. If safety checks are passed, update the EVs.
\end{enumerate}
\textbf{Stopping:}
\begin{enumerate}
\item If $\Delta t>\Delta t_0$ and optimization has converged for current lag time: $\mu_\gamma(\Delta t) < 1.2\mu_\gamma(\Delta t_\infty)$ for $1\le \gamma \le n_\mathrm{ev}$, continue optimization with halved lag time $\Delta t\leftarrow \Delta t/2$.
\item Stop if $\Delta t=\Delta t_0$ and optimization has converged: $\mu_\gamma(\Delta t_0) < \mu_\gamma(\Delta t_\infty)$ for $1\le \gamma \le n_\mathrm{ev}$.
\end{enumerate}

It is advantageous to stop optimization at larger lag times $\Delta t>\Delta t_0$ a bit earlier, i.e., when $\mu_\gamma(\Delta t) < 1.2\mu_\gamma(\Delta t_\infty)$. It, first, speeds up the overall optimization and, second, optimization with smaller lag times continues to improve  $\mu_\gamma(\Delta t)$.

\begin{figure}[htbp]
	\centering
	\includegraphics[width=.5\linewidth]{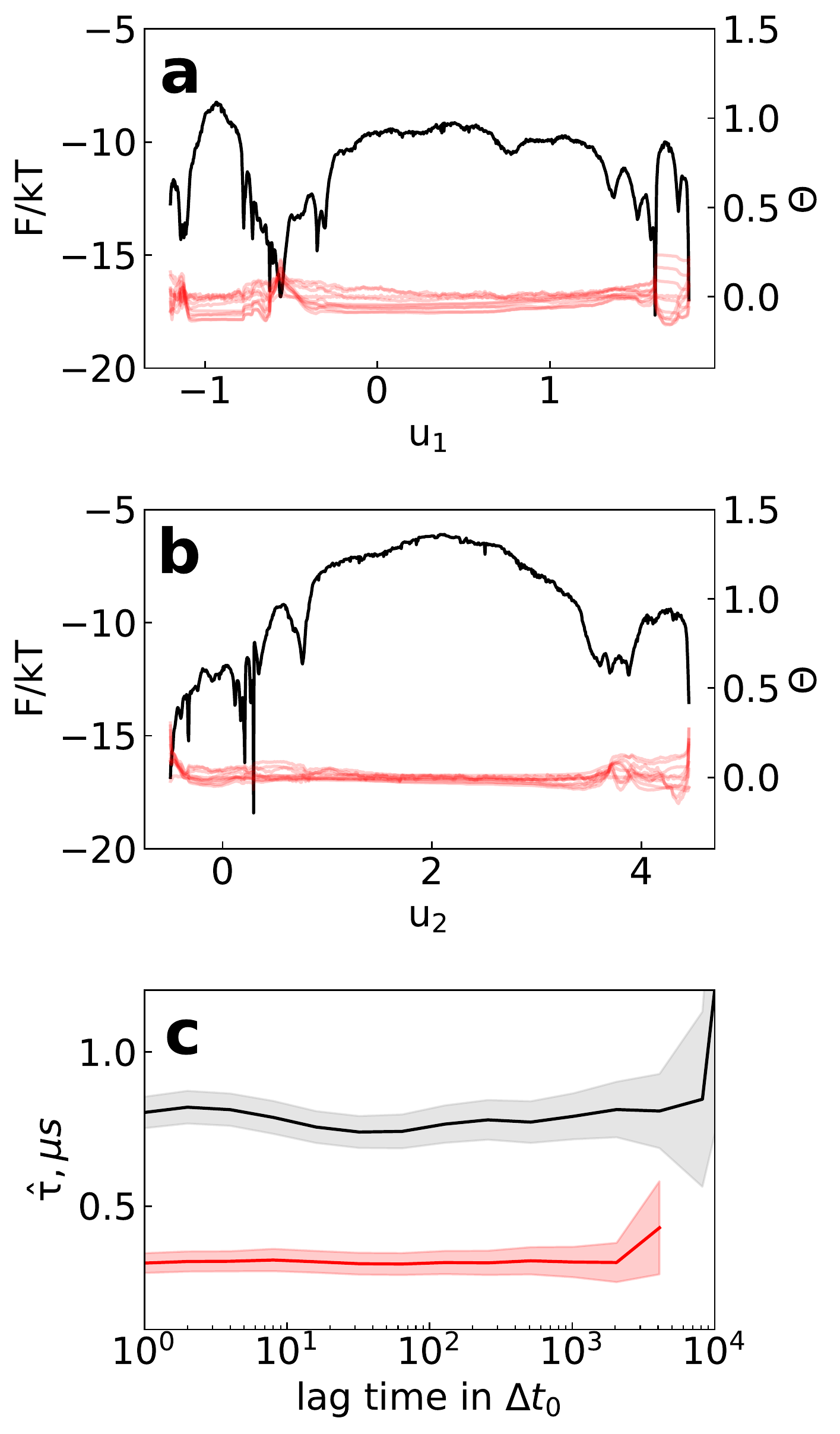}
	\caption{Adaptive non-parametric optimization of first two eigenvectors $u_1$ and $u_2$ of HP35. {\bf a}) Free energy (black) and optimality criterion (red) as functions of $u_1$. {\bf b}) Those as as functions of $u_2$; {\bf c}) Implied timescale $\hat{\tau}$ for $u_1$ (black) and $u_2$ (red) as functions of lag time $\Delta t$; uncertainties (shaded areas) were computed with bootstrap. }
	\label{fig:adapt}
\end{figure}

Fig. \ref{fig:adapt} shows application of the adaptive approach to  determine the first two EVs for the HP35 trajectory. Fig. \ref{fig:adapt}a shows that  $\Theta(x,\Delta t)$ is much closer to zero (bounded by $\pm 0.2$) compared to Fig. \ref{fig:non}c, indicating that $u_1$ is now better approximates the EV. The FEP $F(u_1)$ also shows more structure in the minima. This additional structure disappeared on Fig. \ref{fig:non}a because the regions were not sufficiently optimized. The second EV similarly has $\Theta(x,\Delta t)$ close to zero (Fig. \ref{fig:adapt}b). The implied timescales are accurate starting from the shortest lag time of $0.2$ ns (Fig. \ref{fig:adapt}c).

Note, that it is difficult to compare free energy barriers along different EVs $u_1$ and $u_2$ directly. First, the correspondence between the barriers can be elucidated only by considering the free energy surface as a function of both EVs (see below); for example barrier around $u_1\sim -1$ corresponds to that around $u_2\sim 2$. Second, different EVs provide different, highly nonlinear projections of the configuration space; regions separated on one EV can overlap on another.  

\begin{figure}[htbp]
	\centering
	\includegraphics[width=.5\linewidth]{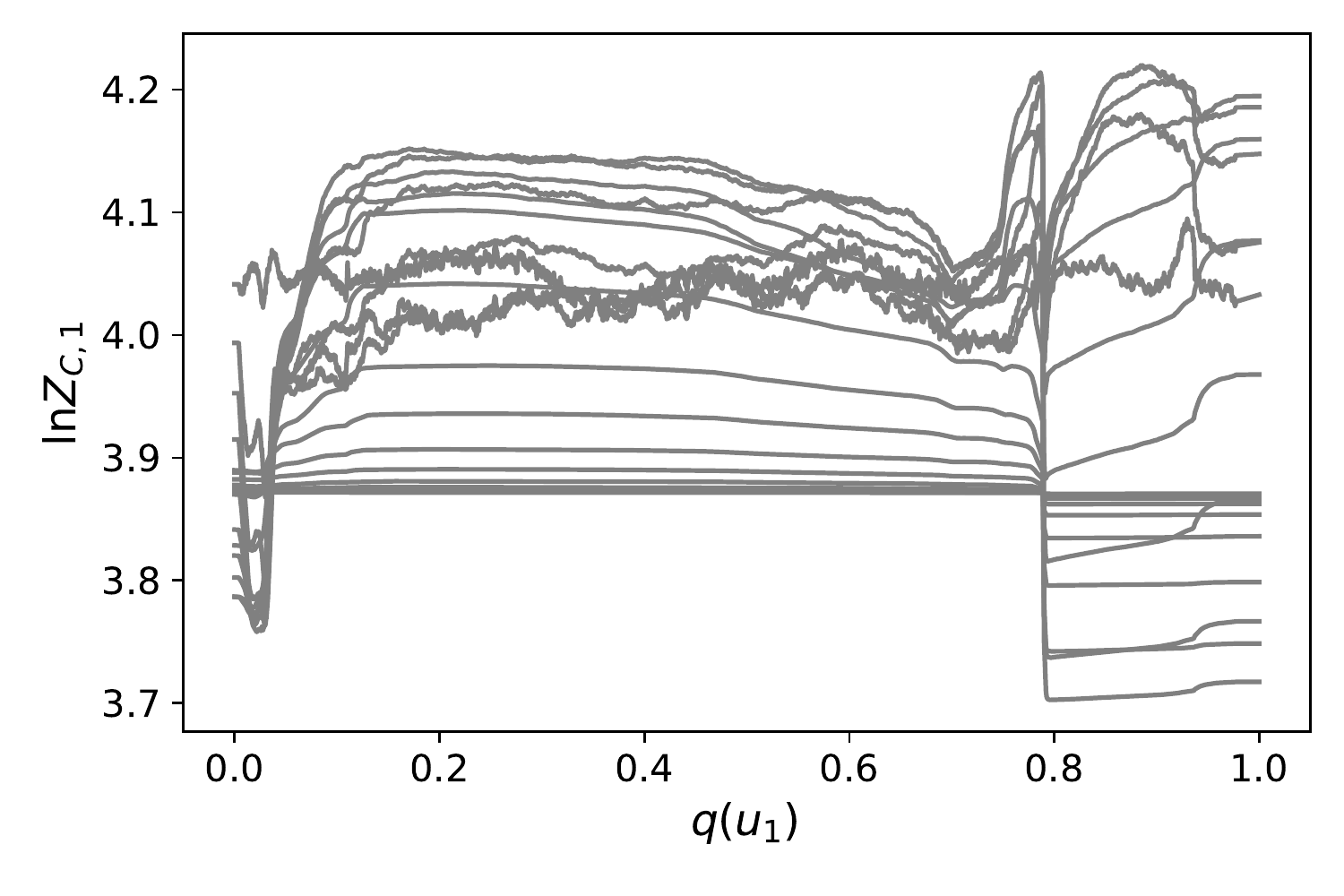}
	\caption{Committor optimality/validation criterion \cite{krivov_reaction_2013} applied to $u_1$. $u_1$ is first transformed to $q(u_1)$, committor as a function of $u_1$. $Z_{C,1}(x,\Delta t)$ along $q(u_1)$ are computed for $\Delta t=1,2,...2^{20}$. Deviations of $\ln Z_{C,1}(x,\Delta t)$ from a constant a bounded by $\pm 0.3$, which translates to relative error around 30\% in estimation of kinetic properties.}
	\label{fig:zc1}
\end{figure}

How accurately do the FEPs on Fig. \ref{fig:adapt} describe the kinetics? For example, the FEP along the committor can be used to compute \textit{exactly} such important properties of kinetics as the equilibrium flux, the mean first passage times, and the mean transition path times between any two regions on the committor \cite{krivov_protein_2018}. Exactly here means that these quantities computed from the one-dimensional diffusion model are equal to that computed directly from the multidimensional trajectory. It, thus, can be used to obtain \textit{direct} accurate estimates of, e.g., free energy barriers and pre-exponential factors \cite{krivov_protein_2018}. The accuracy is limited only by the accuracy of the determined committor. An EV, while being different, could be quite close to the committor between the boundary minima, especially around the transition state (TS) region \cite{berezhkovskii_ensemble_2004}. It can be used to compute the properties approximately. The relative error could be roughly estimated by applying the committor optimality/validation criterion \cite{krivov_reaction_2013} $Z_{C,1}$ to the EV time-series (Fig. \ref{fig:zc1}) and for the first EV the error is around $30$\%. For example, taking boundaries along $u_1$ at $A=-0.565$ and $B=1.8$ (at local minima on Fig \ref{fig:adapt}a) one obtains the following estimates with the diffusive model \cite{krivov_protein_2018}:  
$N_{AB}=58$, mfpt$_{AB}=3974$ ns, mfpt$_{BA}=1194$ ns, mtpt$_{AB}=227$ ns, and directly from trajectory:
$N_{AB}=49$, mfpt$_{AB}=4787$ ns, mfpt$_{BA}=1330$ ns, mtpt$_{AB}=323$ ns; here $N_{AB}$ is the number of transition from A to B, or B to A, mfpt$_{AB}$ is the mean first passage time from A to B, mtpt$_{AB}$ is the mean transition path time between A and B.  For boundaries at $A=-0.565$ and $B=1.607$ estimates from diffusive model are
$N_{AB}=75$, mfpt$_{AB}=2915$ ns, mfpt$_{AB}=1111$ ns, mtpt$_{AB}=61$ ns and directly from trajectory $N_{AB}=77$, mfpt$_{AB}=2790$ ns, mfpt$_{AB}=1102$ ns, mtpt$_{AB}=93$ ns.

\section{Protein folding landscapes and dynamics.}
Using $F(u_i)$ (Fig. \ref{fig:adapt}) for the analysis and description of the dynamics is not very convenient as the diffusion coefficient varies significantly along the EVs \cite{krivov_protein_2018}. It is more convenient to use a ``natural'' coordinate, which we denote as $\tilde{u}_i$, where the diffusion coefficient is constant  $D(\tilde{u}_i)=1$. It is related to $u_i$ by the following monotonous transformation $d\tilde{u}_i/du_i=D(u_i)^{-1/2}$\cite{krivov_diffusive_2008}. 

\begin{figure}[htbp]
	\centering
	\includegraphics[width=.5\linewidth]{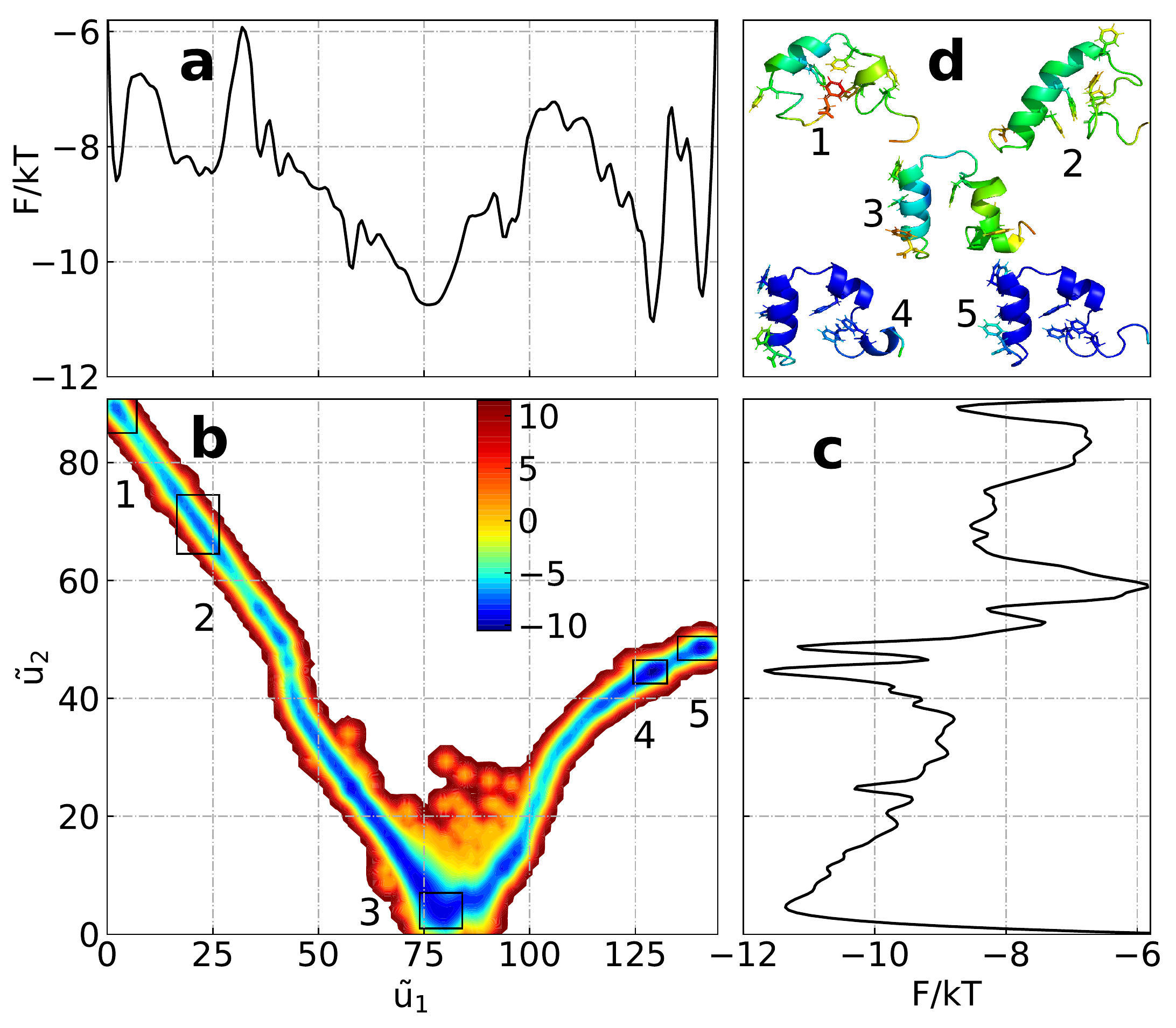}
	\caption{Free energy landscapes of HP35 double mutant: {\bf a}) $F(\tilde{u}_1)$, {\bf c}) $F(\tilde{u}_2)$, and {\bf b}) $F(\tilde{u}_1,\tilde{u}_2)$; the color bar shows $F/kT$. {\bf d}) shows representative structures for the rectangular regions around the free energy minima on {\bf b}); colors code the root-mean-square fluctuations of atomic positions around the average structure from 0.5 \AA{} (blue) to 13 \AA{} (red).}
	\label{fig:2f4k}
\end{figure}

The FEP along the first EV $F(\tilde{u}_1)$ (Fig. \ref{fig:2f4k}a) is in agreement with $F(\tilde{q})$, the FEP along the optimal folding coordinate - the committor between the denatured and the native states; here $\tilde{q}$ denotes the committor monotonously transformed to a natural coordinate \cite{krivov_protein_2018}. 
$F(\tilde{u}_1)$ and $F(\tilde{q})$, in particular, both have minima 3, 4 and 5 and the folding barrier of $\sim 3.5$ kT, confirming that the approach works. There are however also important differences: $F(\tilde{q})$ does not show minima 1 and 2 and minima 4 and 5 are in the opposite order. This is due to the employed definition of the boundary denatured and native states for the committor, defined as structures that have the $C_\alpha$ root-mean-squared-deviation (rmsd) from the native structure smaller than $0.5$ \AA{}  and greater than 10.5 \AA{}, respectively. Using the native minimum (4) with the smallest rmsd as the boundary state forces it to be the rightmost minimum on $F(\tilde{q})$, while $F(\tilde{u}_1)$ reveals that kinetically $5$ is the rightmost minimum. Minima 1, 2 and 3 all have very similar projections on the rmsd, and the boundary state with large rmsd is equally connected to all of them, preventing their separation along $\tilde{q}$. This illustrates that proper definition of boundary states for committor is a difficult problem. As even such a natural approach as using the rmsd leads to inaccuracies. The problem is likely to be more severe for more complex cases, e.g., intrinsically disordered proteins, allosteric transitions, etc, which could be treated by the proposed approach.

Once constructed, the landscapes (Fig. \ref{fig:2f4k}) can be postprocessed to obtain descriptions of minima, TSs, pathways in terms of easy-interpretable coordinates, e.g., dihedral angles, distances \cite{brandt_machine_2018}, or secondary and tertiary structures. For example, since we can easily identify structures that belong to every region of the FES, a supervised machine learning model can be trained
to assign these structures to these regions. It will make the model to learn to identify the most important molecular coordinates, e.g., inter-atom distances or dihedral angles, that discriminate these states \cite{brandt_machine_2018}. Or, more generally, one can consider a standard machine learning regression problem of approximating the determined EVs coordinates $u_1$ and $u_2$, by a function of e.g., selected collective variables or inter-atom distances or dihedral angles. The regression problem is simpler then the original problem of accurate determination of the slowest EVs. Note, that it is, probably, easier to approximate $\tilde{u}_1$ and $\tilde{u}_2$, where TS and minima have similar scales.

Here we analyze the FES in terms of tertiary structures. For every free energy minimum we find the geometric average of all the structures in the minimum. Each structure is optimally superposed on the first trajectory structure of the minimum. A structure from the trajectory closest to the geometric average is found and is considered as a representative structure for the minimum. The process is repeated a few iterations with all the structures superposing on the representative structure, until the latter stops changing. The root-mean-square fluctuations (rmsf) for every residue is computed as the square root of the mean squared distances of all the atoms of the residue between the representative structure and all the superposed structures. Cartoon pictures of representative structures, colored according to the rmsf, from 0.5 \AA{} (blue) to 13 \AA{} (red) are shown on Fig. \ref{fig:2f4k}d.

In minimum 3 the protein is almost folded: all three helices are formed with a relatively high propensity and are all at the right positions. The hydrophobic core is not formed and the structure is rather flexible. In the native minimum (4) the folding is completed by forming the hydrophobic core and making the structure stable. Near-native minimum 5 has first and third helices partially unraveled \cite{beauchamp_simple_2012}. In minimum 3 residues 18-24 form a turn, connecting second and third helices, whereas in minima 1 and 2, they form a helix with $>90$ \% propensity. It leads to the possibility of the second and third helices forming a single long helix in minimum 2 and a longer second helix in 1.

The two-dimensional FES $F(\tilde{u}_1,\tilde{u}_2)$ can be used to find the correspondence among the minima on the FEPs and see the evidence of parallel pathways. In particular, the FES for HP35 has an L-like shape and shows no evidence of parallel pathways. The one-dimensional FEPs, i.e., $F(\tilde{u}_1)$, on the other hand, are better suited for the quantitative analysis of the dynamics, like determining free energies of TSs and minima, free energy barriers and pre-exponential factors, computing rates, mean first passage times, etc. 

\begin{figure}[htbp]
\includegraphics[width=.5\linewidth]{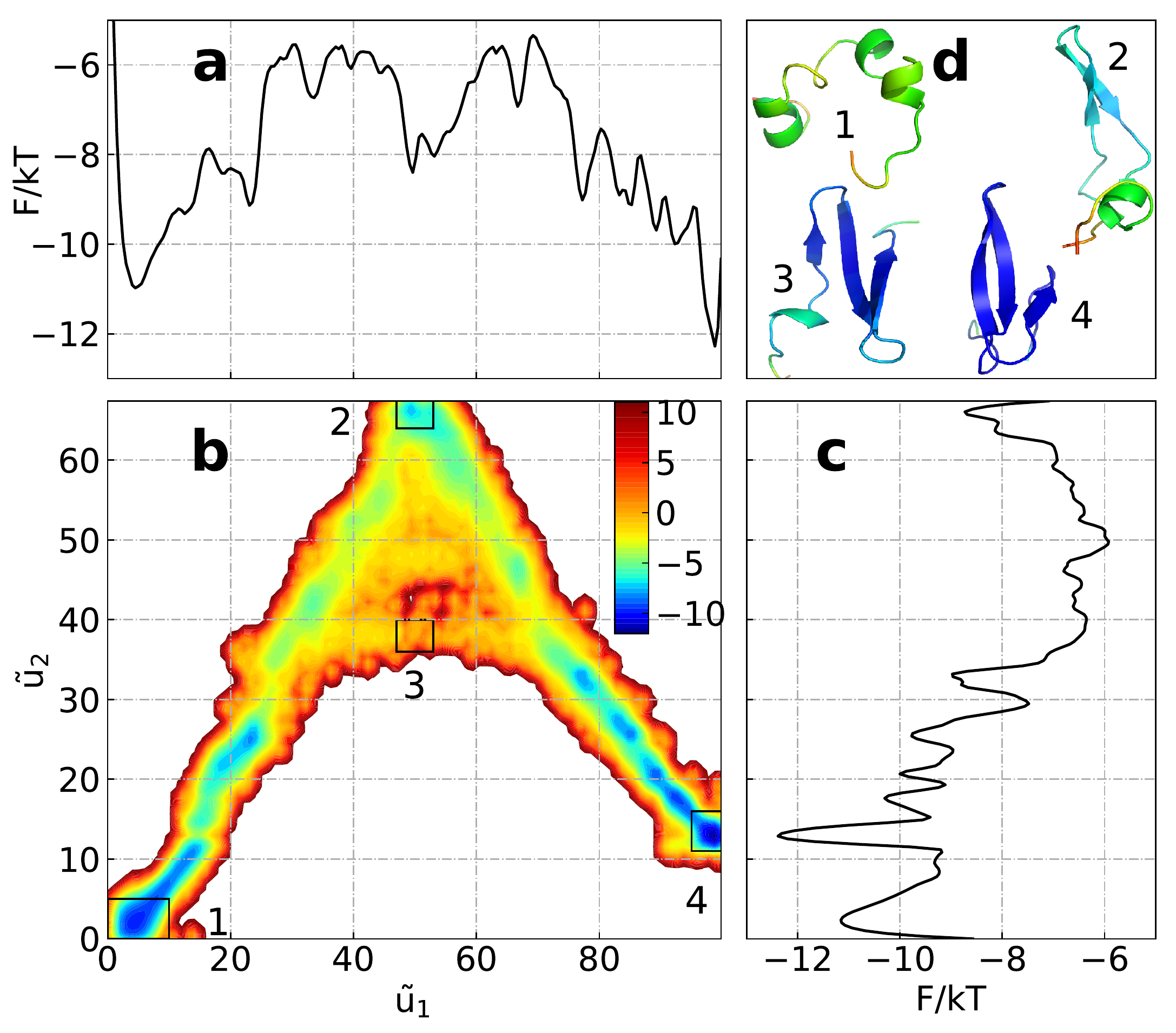}
\caption{Free energy landscapes of FIP35: notation as in Fig. \ref{fig:2f4k}. }
\label{fig:ww}
\end{figure}

We have also applied the approach the FIP35 protein trajectory (Fig. \ref{fig:ww}) \cite{shaw_atomic-level_2010}. The EV validation test $\Theta(x,\Delta t)$ was bounded by $\pm 0.2$ for both EVs. This trajectory has only 15 folding-unfolding events, which illustrates that the approach can analyze systems with very limited sampling.  $F(\tilde{u}_1)$ shows two minima with an intermediate state in agreement with other studies \cite{krivov_free_2011, boninsegna_investigating_2015}. The two-dimensional FES has an A-like shape and shows the evidence of two parallel pathways, i.e., protein folds from 1 to 4 via 2 or 3. The representative structure of 2 has the first hairpin formed, while that of 3 has the second hairpin formed to a much larger degree. Surprisingly, region 3 is a TS rather than an intermediate state. It probably explains why this pathway is much less populated. It might be difficult to detect this pathway using MSMs. The intermediate TS is much less populated, thus, a rather large clustering size could be required to have a representative statistics. However, a large clustering size will make it more likely that points from the TS are assigned to free energy minima, which are much more populated.

\section{Concluding Discussion}
We have described a blind approach for the determination of the slowest relaxation eigenvectors from an equilibrium trajectory. The approach determined the first and second slowest eigenvectors for the HP35 and FIP35 proteins with high spatio-temporal accuracy, as validated by the stringent criterion at the shortest lag time of $0.2$ ns. In contrast to alternative (parametric) approaches, which require approximating functions with many parameters and extensive expertise with the system, the approach directly determines eigenvectors time-series and uses no system specific information. The optimality criterion is another important ingredient of the approach, which makes the uniform optimization possible. The approach can be used in cases when one does not want to introduce any bias in the analysis, e.g., due to employed approximating functions, or one does not have good approximating functions. It can also be used aposteriory to check if possible bias in the analysis has altered the results. As the HP35 example illustrates, even a seemingly innocent and natural choice of boundary states can hide the inherent complexity of the landscape. 

The approach was illustrated by analyzing long equilibrium trajectories, i.e., state-of-the-art atomistic protein folding simulations  \cite{shaw_atomic-level_2010, lindorff-larsen_how_2011}. However, generating such trajectories by brute-force molecular dynamics is very computationally demanding. A number of advanced sampling methods have been suggested to alleviate the sampling problem, e.g, umbrella sampling \cite{torrie_nonphysical_1977, souaille_umbrella_2001}, steered-MD \cite{isralewitz_steered_2001}, replica-exchange \cite{sugita_replica_1999, fukunishi_replica_2002}, meta-dynamics \cite{barducci_metadynamics_2011}. If an enhanced sampling method generates unbiased, equilibrium sampling, possibly consisting of many trajectories, e.g., the trajectories of the base replica of the recently suggested REHT method \cite{appadurai_reht_2021}, they can be analyzed by the approach directly. One just needs to extend the summation to all the trajectories in Eq. \ref{v:a}. For other approaches, which can be used to determine equilibrium probabilities, however perturb the natural dynamics of the system, the equilibrium sampling needs to be generated first. It can be done, e.g.,  by starting many trajectories with natural, unbiased dynamics from the obtained configurations with equilibrium probabilities. It is possible to extend the developed non-parametric approaches to non-equilibrium sampling, which is generated, for example, by adaptive sampling methods. Mainly, it requires the change of the optimization functional and correspondingly the equations for optimal parameters of the variation (Eq. \ref{v:a} here) and is discussed elsewhere \cite{krivov_npne_2021}.

Here, we consider EV as a function of a trajectory, rather than a function of configuration space. In principle, it is possible to record all the parameters of the transformations and collective variables during EV optimization (training) and apply them later, in the same order, to new (test) data. That would make the determined EV a function of the configuration space. Here, however, we did not record the transformations, and computed the EV only for the configurations along a trajectory. 

It is instructive to compare the proposed approach with alternative approaches. Diffusion maps \cite{coifman_diffusion_2006}, Laplacian eigenmaps \cite{belkin_laplacian_2003}  and Isomap \cite{tenenbaum_global_2000} are non-parametric generic dimensionality reduction methods. The main difference between these approaches and the proposed approach is that the former analyse a model of the dynamics, while the later - the actual dynamics. For example, the diffusion maps and Laplacian eigenmaps effectively define transition matrix between the configurations as the heat (diffusion) kernel according to the distance between them. It can be said, loosely, that these methods perform dimensionality reduction with a focus on preserving the properties (proximity) of a given configuration space. However, it is well known that the geometric proximity is a poor indicator of kinetics proximity. Configurations which are close geometrically can be separated by high barriers, while motions along the low energy normal modes (i.e., low barriers) are generally associated with large conformational changes.

A large collection of parametric approaches, e.g., 
tICA \cite{schwantes_improvements_2013,perez_identification_2013}, VAMP \cite{wu_vamp_2017}, EDMD \cite{wu_vamp_2017, williams_data-driven_2015} aims to approximate the slowest eigenvectors by multy-parametric functions, for example, a linear combination of collective variables or a neural network. Their major weakness is that their performance is limited by the choice of the employed functional forms and the input/collective variables. Since  finding, e.g., an optimal architecture of a neural network or informative collective variables are difficult tasks. While intuition can help to solve these problems for low-dimensional model systems, the difficulty in the case of complex realistic systems becomes apparent, when one realizes that such a function should be able to accurately project a few million snapshots of a very high-dimensional trajectory. In particular, it implies an extensive knowledge of the system, and that an acceptable solution is likely to be system specific. 

The proposed method is non-parametric and can approximate any EV with high accuracy. While each iteration may depend on the exact choice of the family of collective variables/molecular descriptors/features, the final EVs do not, since they provide optimum to a (non-parametric) target functional, when the optimization converges. We assume that the employed input variables contain all the information about the dynamics of interest. For the analysis of biomolecular simulations one can suggest the inter-atom distances as the standard sets of input variables. The iterative optimization of EVs, using these standard input variables, is a more generic and a more efficient approach than custom design of multi-parametric functions. As Fig. \ref{fig:spur}a shows, a few thousands iterations can provide a rather good approximation to an EV with the corresponding eigenvalue within a small factor from the exact value. The determined EVs pass a stringent validation test at a very short time-scales of the trajectory sampling interval. It means that the obtained EVs time-series are more accurate than those obtained with alternative approaches. They provide a higher temporal resolution in the description of the dynamics. Much shorter lag time also means that much shorter trajectories are required for the simple strategy of exascale simulations and thus a much larger possible speedup over direct, brute-force simulations.

\section*{Acknowledgments}
I am grateful to David Shaw and his coworkers for making the folding trajectories available.

\section{Appendix}

\subsection{Adaptive non-parametric optimization of eigenvectors}
The simple, non-adaptive algorithm, optimizes EVs in a non-uniform way analogous to the committor case. It is easier to optimize free energy barriers than minima. To perform optimization in a uniform way one needs first to detect sub-optimal regions of EVs and focus optimization on them. To detect the most suboptimal regions for current lag time $\Delta t$, we first find a longer lag time $\Delta t_1$, which exhibits the most nonuniformness in the distance between $\Theta(x,\Delta t_1)$ and $\Theta(x,\Delta t)$: 
\begin{equation}
	\Delta t_1=\arg \sup_{t_i} [\max_x \Delta \Theta(x,\Delta t_i,\Delta t) - \min_x \Delta \Theta(x,\Delta t_i,\Delta t)],
\end{equation}
here $\Delta \Theta(x,\Delta t_i,\Delta t)=\Theta(x,\Delta t_i) - \Theta(x,\Delta t)$. Then, the relative degree of suboptimality of region around $x$ is defined as 
\begin{equation}
	s(x)=\exp[\Delta \Theta(x,\Delta t_1,\Delta t) -\max_x \Delta \Theta(x,\Delta t_1,\Delta t)]
\end{equation}
It takes maximal value of $1$ for the most suboptimal part where the difference between $\Theta(x,\Delta t_1)$ and $\Theta(x,\Delta t)$ is maximal. To focus optimization on such suboptimal regions we make $p_\mathrm{fix}$ position dependent, large for optimal regions and small for suboptimal regions. Consequently, the optimization is more focused on less optimized regions, because they have a smaller number of fixed points and are less constraint. For example, an extremely over-optimized region might have $p_\mathrm{fix}=1$, i.e., all the points fixed and thus it will not be optimized at all.
Here we used 
\begin{equation}
	p_\mathrm{fix}(x)=\min[1,p_\mathrm{fix}\times s(x)^{-10}]
\end{equation}
Before every iteration, the $p_\mathrm{fix}(x)$ values are computed for active (k-th) eigenvector, and are used to select fixed points. Namely, a point at time moment $i \Delta t$, that has eigenvector coordinate  $u_\beta(i\Delta t)$ is selected to be fixed with probability $p_\mathrm{fix}(u_\beta(i\Delta t))$.


\begin{mcitethebibliography}{41}
	\providecommand*\natexlab[1]{#1}
	\providecommand*\mciteSetBstSublistMode[1]{}
	\providecommand*\mciteSetBstMaxWidthForm[2]{}
	\providecommand*\mciteBstWouldAddEndPuncttrue
	{\def\EndOfBibitem{\unskip.}}
	\providecommand*\mciteBstWouldAddEndPunctfalse
	{\let\EndOfBibitem\relax}
	\providecommand*\mciteSetBstMidEndSepPunct[3]{}
	\providecommand*\mciteSetBstSublistLabelBeginEnd[3]{}
	\providecommand*\EndOfBibitem{}
	\mciteSetBstSublistMode{f}
	\mciteSetBstMaxWidthForm{subitem}{(\alph{mcitesubitemcount})}
	\mciteSetBstSublistLabelBeginEnd
	{\mcitemaxwidthsubitemform\space}
	{\relax}
	{\relax}
	
	\bibitem[Shaw \latin{et~al.}(2010)Shaw, Maragakis, Lindorff-Larsen, Piana,
	Dror, Eastwood, Bank, Jumper, Salmon, Shan, and
	Wriggers]{shaw_atomic-level_2010}
	Shaw,~D.~E.; Maragakis,~P.; Lindorff-Larsen,~K.; Piana,~S.; Dror,~R.~O.;
	Eastwood,~M.~P.; Bank,~J.~A.; Jumper,~J.~M.; Salmon,~J.~K.; Shan,~Y.;
	Wriggers,~W. Atomic-{Level} {Characterization} of the {Structural} {Dynamics}
	of {Proteins}. \emph{Science} \textbf{2010}, \emph{330}, 341--346\relax
	\mciteBstWouldAddEndPuncttrue
	\mciteSetBstMidEndSepPunct{\mcitedefaultmidpunct}
	{\mcitedefaultendpunct}{\mcitedefaultseppunct}\relax
	\EndOfBibitem
	\bibitem[Lindorff-Larsen \latin{et~al.}(2011)Lindorff-Larsen, Piana, Dror, and
	Shaw]{lindorff-larsen_how_2011}
	Lindorff-Larsen,~K.; Piana,~S.; Dror,~R.~O.; Shaw,~D.~E. How {Fast}-{Folding}
	{Proteins} {Fold}. \emph{Science} \textbf{2011}, \emph{334}, 517--520\relax
	\mciteBstWouldAddEndPuncttrue
	\mciteSetBstMidEndSepPunct{\mcitedefaultmidpunct}
	{\mcitedefaultendpunct}{\mcitedefaultseppunct}\relax
	\EndOfBibitem
	\bibitem[Freddolino \latin{et~al.}(2010)Freddolino, Harrison, Liu, and
	Schulten]{freddolino_challenges_2010}
	Freddolino,~P.~L.; Harrison,~C.~B.; Liu,~Y.; Schulten,~K. Challenges in
	protein-folding simulations. \emph{Nat Phys} \textbf{2010}, \emph{6},
	751--758\relax
	\mciteBstWouldAddEndPuncttrue
	\mciteSetBstMidEndSepPunct{\mcitedefaultmidpunct}
	{\mcitedefaultendpunct}{\mcitedefaultseppunct}\relax
	\EndOfBibitem
	\bibitem[Schwantes and Pande(2015)Schwantes, and
	Pande]{schwantes_modeling_2015}
	Schwantes,~C.~R.; Pande,~V.~S. Modeling {Molecular} {Kinetics} with {tICA} and
	the {Kernel} {Trick}. \emph{J. Chem. Theory Comput.} \textbf{2015},
	\emph{11}, 600--608\relax
	\mciteBstWouldAddEndPuncttrue
	\mciteSetBstMidEndSepPunct{\mcitedefaultmidpunct}
	{\mcitedefaultendpunct}{\mcitedefaultseppunct}\relax
	\EndOfBibitem
	\bibitem[Banushkina and Krivov(2016)Banushkina, and
	Krivov]{banushkina_optimal_2016}
	Banushkina,~P.~V.; Krivov,~S.~V. Optimal reaction coordinates. \emph{WIREs
		Comput Mol Sci} \textbf{2016}, \emph{6}, 748--763\relax
	\mciteBstWouldAddEndPuncttrue
	\mciteSetBstMidEndSepPunct{\mcitedefaultmidpunct}
	{\mcitedefaultendpunct}{\mcitedefaultseppunct}\relax
	\EndOfBibitem
	\bibitem[No{\'e} and Clementi(2017)No{\'e}, and Clementi]{noe_collective_2017}
	No{\'e},~F.; Clementi,~C. Collective variables for the study of long-time
	kinetics from molecular trajectories: theory and methods. \emph{Curr. Opin.
		Struct. Biol.} \textbf{2017}, \emph{43}, 141--147\relax
	\mciteBstWouldAddEndPuncttrue
	\mciteSetBstMidEndSepPunct{\mcitedefaultmidpunct}
	{\mcitedefaultendpunct}{\mcitedefaultseppunct}\relax
	\EndOfBibitem
	\bibitem[Jung \latin{et~al.}(2019)Jung, Covino, and
	Hummer]{jung_artificial_2019}
	Jung,~H.; Covino,~R.; Hummer,~G. Artificial Intelligence Assists Discovery of
	Reaction Coordinates and Mechanisms from Molecular Dynamics Simulations.
	\textbf{2019}, arXiv: 1901.04595 [physics:chem--ph]\relax
	\mciteBstWouldAddEndPuncttrue
	\mciteSetBstMidEndSepPunct{\mcitedefaultmidpunct}
	{\mcitedefaultendpunct}{\mcitedefaultseppunct}\relax
	\EndOfBibitem
	\bibitem[Peters(2015)]{peters_common_2015}
	Peters,~B. Common {Features} of {Extraordinary} {Rate} {Theories}. \emph{J.
		Phys. Chem. B} \textbf{2015}, \emph{119}, 6349--6356\relax
	\mciteBstWouldAddEndPuncttrue
	\mciteSetBstMidEndSepPunct{\mcitedefaultmidpunct}
	{\mcitedefaultendpunct}{\mcitedefaultseppunct}\relax
	\EndOfBibitem
	\bibitem[Peters(2016)]{peters_reaction_2016}
	Peters,~B. Reaction {Coordinates} and {Mechanistic} {Hypothesis} {Tests}.
	\emph{Ann. Rev. Phys. Chem.} \textbf{2016}, \emph{67}, 669--690\relax
	\mciteBstWouldAddEndPuncttrue
	\mciteSetBstMidEndSepPunct{\mcitedefaultmidpunct}
	{\mcitedefaultendpunct}{\mcitedefaultseppunct}\relax
	\EndOfBibitem
	\bibitem[Krivov(2018)]{krivov_protein_2018}
	Krivov,~S.~V. Protein Folding Free Energy Landscape along the Committor - the
	Optimal Folding Coordinate. \emph{J. Chem. Theory Comput.} \textbf{2018},
	\emph{14}, 3418--3427\relax
	\mciteBstWouldAddEndPuncttrue
	\mciteSetBstMidEndSepPunct{\mcitedefaultmidpunct}
	{\mcitedefaultendpunct}{\mcitedefaultseppunct}\relax
	\EndOfBibitem
	\bibitem[Shuler(1959)]{shuler_relaxation_1959}
	Shuler,~K.~E. Relaxation Processes in Multistate Systems. \emph{The Physics of
		Fluids} \textbf{1959}, \emph{2}, 442--448\relax
	\mciteBstWouldAddEndPuncttrue
	\mciteSetBstMidEndSepPunct{\mcitedefaultmidpunct}
	{\mcitedefaultendpunct}{\mcitedefaultseppunct}\relax
	\EndOfBibitem
	\bibitem[McGibbon \latin{et~al.}(2017)McGibbon, Husic, and
	Pande]{mcgibbon_identification_2017}
	McGibbon,~R.~T.; Husic,~B.~E.; Pande,~V.~S. Identification of simple reaction
	coordinates from complex dynamics. \emph{J Chem Phys} \textbf{2017},
	\emph{146}, 044109\relax
	\mciteBstWouldAddEndPuncttrue
	\mciteSetBstMidEndSepPunct{\mcitedefaultmidpunct}
	{\mcitedefaultendpunct}{\mcitedefaultseppunct}\relax
	\EndOfBibitem
	\bibitem[Schwantes and Pande(2013)Schwantes, and
	Pande]{schwantes_improvements_2013}
	Schwantes,~C.~R.; Pande,~V.~S. Improvements in Markov State Model Construction
	Reveal Many Non-Native Interactions in the Folding of NTL9. \emph{J. Chem.
		Theory Comput.} \textbf{2013}, \emph{9}, 2000--2009\relax
	\mciteBstWouldAddEndPuncttrue
	\mciteSetBstMidEndSepPunct{\mcitedefaultmidpunct}
	{\mcitedefaultendpunct}{\mcitedefaultseppunct}\relax
	\EndOfBibitem
	\bibitem[Pérez-Hernández \latin{et~al.}(2013)Pérez-Hernández, Paul,
	Giorgino, De~Fabritiis, and Noé]{perez_identification_2013}
	Pérez-Hernández,~G.; Paul,~F.; Giorgino,~T.; De~Fabritiis,~G.; Noé,~F.
	Identification of slow molecular order parameters for Markov model
	construction. \emph{J. Chem. Phys.} \textbf{2013}, \emph{139}, 015102\relax
	\mciteBstWouldAddEndPuncttrue
	\mciteSetBstMidEndSepPunct{\mcitedefaultmidpunct}
	{\mcitedefaultendpunct}{\mcitedefaultseppunct}\relax
	\EndOfBibitem
	\bibitem[Wan and Voelz(2020)Wan, and Voelz]{wan_adaptive_2020}
	Wan,~H.; Voelz,~V.~A. Adaptive Markov state model estimation using short
	reseeding trajectories. \emph{J. Chem. Phys.} \textbf{2020}, \emph{152},
	024103\relax
	\mciteBstWouldAddEndPuncttrue
	\mciteSetBstMidEndSepPunct{\mcitedefaultmidpunct}
	{\mcitedefaultendpunct}{\mcitedefaultseppunct}\relax
	\EndOfBibitem
	\bibitem[Hern\'andez \latin{et~al.}(2018)Hern\'andez, Wayment-Steele, Sultan,
	Husic, and Pande]{hernandez_variational_2018}
	Hern\'andez,~C.~X.; Wayment-Steele,~H.~K.; Sultan,~M.~M.; Husic,~B.~E.;
	Pande,~V.~S. Variational encoding of complex dynamics. \emph{Phys. Rev. E}
	\textbf{2018}, \emph{97}, 062412\relax
	\mciteBstWouldAddEndPuncttrue
	\mciteSetBstMidEndSepPunct{\mcitedefaultmidpunct}
	{\mcitedefaultendpunct}{\mcitedefaultseppunct}\relax
	\EndOfBibitem
	\bibitem[Mardt \latin{et~al.}(2018)Mardt, Pasquali, Wu, and
	Noé]{mardt_vampnets_2020}
	Mardt,~A.; Pasquali,~L.; Wu,~H.; Noé,~F. VAMPnets for deep learning of
	molecular kinetics. \emph{Nature Communications} \textbf{2018}, \emph{9},
	5\relax
	\mciteBstWouldAddEndPuncttrue
	\mciteSetBstMidEndSepPunct{\mcitedefaultmidpunct}
	{\mcitedefaultendpunct}{\mcitedefaultseppunct}\relax
	\EndOfBibitem
	\bibitem[Banushkina and Krivov(2015)Banushkina, and
	Krivov]{banushkina_nonparametric_2015}
	Banushkina,~P.~V.; Krivov,~S.~V. Nonparametric variational optimization of
	reaction coordinates. \emph{J. Chem. Phys.} \textbf{2015}, \emph{143},
	184108\relax
	\mciteBstWouldAddEndPuncttrue
	\mciteSetBstMidEndSepPunct{\mcitedefaultmidpunct}
	{\mcitedefaultendpunct}{\mcitedefaultseppunct}\relax
	\EndOfBibitem
	\bibitem[Piana \latin{et~al.}(2012)Piana, Lindorff-Larsen, and
	Shaw]{piana_protein_2012}
	Piana,~S.; Lindorff-Larsen,~K.; Shaw,~D.~E. Protein folding kinetics and
	thermodynamics from atomistic simulation. \emph{PNAS} \textbf{2012},
	\emph{109}, 17845--17850\relax
	\mciteBstWouldAddEndPuncttrue
	\mciteSetBstMidEndSepPunct{\mcitedefaultmidpunct}
	{\mcitedefaultendpunct}{\mcitedefaultseppunct}\relax
	\EndOfBibitem
	\bibitem[Krivov(2013)]{krivov_reaction_2013}
	Krivov,~S.~V. On {Reaction} {Coordinate} {Optimality}. \emph{J. Chem. Theory
		Comput.} \textbf{2013}, \emph{9}, 135--146\relax
	\mciteBstWouldAddEndPuncttrue
	\mciteSetBstMidEndSepPunct{\mcitedefaultmidpunct}
	{\mcitedefaultendpunct}{\mcitedefaultseppunct}\relax
	\EndOfBibitem
	\bibitem[Krivov(2020)]{CFEPs}
	Krivov,~S. CFEPs. \url{https://github.com/krivovsv/CFEPs}, 2020\relax
	\mciteBstWouldAddEndPuncttrue
	\mciteSetBstMidEndSepPunct{\mcitedefaultmidpunct}
	{\mcitedefaultendpunct}{\mcitedefaultseppunct}\relax
	\EndOfBibitem
	\bibitem[Berezhkovskii and Szabo(2004)Berezhkovskii, and
	Szabo]{berezhkovskii_ensemble_2004}
	Berezhkovskii,~A.; Szabo,~A. Ensemble of transition states for two-state
	protein folding from the eigenvectors of rate matrices. \emph{J. Chem. Phys.}
	\textbf{2004}, \emph{121}, 9186--9187\relax
	\mciteBstWouldAddEndPuncttrue
	\mciteSetBstMidEndSepPunct{\mcitedefaultmidpunct}
	{\mcitedefaultendpunct}{\mcitedefaultseppunct}\relax
	\EndOfBibitem
	\bibitem[Krivov and Karplus(2008)Krivov, and Karplus]{krivov_diffusive_2008}
	Krivov,~S.~V.; Karplus,~M. Diffusive reaction dynamics on invariant free energy
	profiles. \emph{PNAS} \textbf{2008}, \emph{105}, 13841--13846\relax
	\mciteBstWouldAddEndPuncttrue
	\mciteSetBstMidEndSepPunct{\mcitedefaultmidpunct}
	{\mcitedefaultendpunct}{\mcitedefaultseppunct}\relax
	\EndOfBibitem
	\bibitem[Brandt \latin{et~al.}(2018)Brandt, Sittel, Ernst, and
	Stock]{brandt_machine_2018}
	Brandt,~S.; Sittel,~F.; Ernst,~M.; Stock,~G. Machine Learning of Biomolecular
	Reaction Coordinates. \emph{J. Phys. Chem. Lett.} \textbf{2018}, \emph{9},
	2144--2150\relax
	\mciteBstWouldAddEndPuncttrue
	\mciteSetBstMidEndSepPunct{\mcitedefaultmidpunct}
	{\mcitedefaultendpunct}{\mcitedefaultseppunct}\relax
	\EndOfBibitem
	\bibitem[Beauchamp \latin{et~al.}(2012)Beauchamp, McGibbon, Lin, and
	Pande]{beauchamp_simple_2012}
	Beauchamp,~K.~A.; McGibbon,~R.; Lin,~Y.-S.; Pande,~V.~S. Simple few-state
	models reveal hidden complexity in protein folding. \emph{PNAS}
	\textbf{2012}, \emph{109}, 17807--17813\relax
	\mciteBstWouldAddEndPuncttrue
	\mciteSetBstMidEndSepPunct{\mcitedefaultmidpunct}
	{\mcitedefaultendpunct}{\mcitedefaultseppunct}\relax
	\EndOfBibitem
	\bibitem[Krivov(2011)]{krivov_free_2011}
	Krivov,~S.~V. The {Free} {Energy} {Landscape} {Analysis} of {Protein} ({FIP}35)
	{Folding} {Dynamics}. \emph{J. Phys. Chem. B} \textbf{2011}, \emph{115},
	12315--12324\relax
	\mciteBstWouldAddEndPuncttrue
	\mciteSetBstMidEndSepPunct{\mcitedefaultmidpunct}
	{\mcitedefaultendpunct}{\mcitedefaultseppunct}\relax
	\EndOfBibitem
	\bibitem[Boninsegna \latin{et~al.}(2015)Boninsegna, Gobbo, Noe, and
	Clementi]{boninsegna_investigating_2015}
	Boninsegna,~L.; Gobbo,~G.; Noe,~F.; Clementi,~C. Investigating {Molecular}
	{Kinetics} by {Variationally} {Optimized} {Diffusion} {Maps}. \emph{J. Chem.
		Theory Comput.} \textbf{2015}, \emph{11}, 5947--5960\relax
	\mciteBstWouldAddEndPuncttrue
	\mciteSetBstMidEndSepPunct{\mcitedefaultmidpunct}
	{\mcitedefaultendpunct}{\mcitedefaultseppunct}\relax
	\EndOfBibitem
	\bibitem[Torrie and Valleau(1977)Torrie, and Valleau]{torrie_nonphysical_1977}
	Torrie,~G.~M.; Valleau,~J.~P. Nonphysical sampling distributions in {Monte}
	{Carlo} free-energy estimation: {Umbrella} sampling. \emph{J Comput. Phys.}
	\textbf{1977}, \emph{23}, 187--199\relax
	\mciteBstWouldAddEndPuncttrue
	\mciteSetBstMidEndSepPunct{\mcitedefaultmidpunct}
	{\mcitedefaultendpunct}{\mcitedefaultseppunct}\relax
	\EndOfBibitem
	\bibitem[Souaille and Roux(2001)Souaille, and Roux]{souaille_umbrella_2001}
	Souaille,~M.; Roux,~B. Extension to the Weighted Histogram Analysis Method:
	Combining Umbrella Sampling with Free Energy Calculations. \emph{Comput.
		Phys. Commun.} \textbf{2001}, \emph{135}, 40\relax
	\mciteBstWouldAddEndPuncttrue
	\mciteSetBstMidEndSepPunct{\mcitedefaultmidpunct}
	{\mcitedefaultendpunct}{\mcitedefaultseppunct}\relax
	\EndOfBibitem
	\bibitem[Isralewitz \latin{et~al.}(2001)Isralewitz, Baudry, Gullingsrud,
	Kosztin, and Schulten]{isralewitz_steered_2001}
	Isralewitz,~B.; Baudry,~J.; Gullingsrud,~J.; Kosztin,~D.; Schulten,~K. Steered
	Molecular Dynamics Investigations of Protein Function. \emph{J. Mol. Graphics
		Modell.} \textbf{2001}, \emph{19}, 13\relax
	\mciteBstWouldAddEndPuncttrue
	\mciteSetBstMidEndSepPunct{\mcitedefaultmidpunct}
	{\mcitedefaultendpunct}{\mcitedefaultseppunct}\relax
	\EndOfBibitem
	\bibitem[Sugita and Okamoto(1999)Sugita, and Okamoto]{sugita_replica_1999}
	Sugita,~Y.; Okamoto,~Y. Replica-exchange molecular dynamics method for protein
	folding. \emph{Chem. Phys. Lett.} \textbf{1999}, \emph{314}, 141\relax
	\mciteBstWouldAddEndPuncttrue
	\mciteSetBstMidEndSepPunct{\mcitedefaultmidpunct}
	{\mcitedefaultendpunct}{\mcitedefaultseppunct}\relax
	\EndOfBibitem
	\bibitem[Fukunishi \latin{et~al.}(2002)Fukunishi, Watanabe, and
	Takada]{fukunishi_replica_2002}
	Fukunishi,~H.; Watanabe,~O.; Takada,~S. On the Hamiltonian replica exchange
	method for efficient sampling of biomolecular systems: Application to protein
	structure prediction. \emph{J. Chem. Phys.} \textbf{2002}, \emph{116},
	9058\relax
	\mciteBstWouldAddEndPuncttrue
	\mciteSetBstMidEndSepPunct{\mcitedefaultmidpunct}
	{\mcitedefaultendpunct}{\mcitedefaultseppunct}\relax
	\EndOfBibitem
	\bibitem[Barducci \latin{et~al.}(2011)Barducci, Bonomi, and
	Parrinello]{barducci_metadynamics_2011}
	Barducci,~A.; Bonomi,~M.; Parrinello,~M. Metadynamics. \emph{Wiley Interdiscip.
		Rev. Comput. Mol. Sci.} \textbf{2011}, \emph{1}, 826\relax
	\mciteBstWouldAddEndPuncttrue
	\mciteSetBstMidEndSepPunct{\mcitedefaultmidpunct}
	{\mcitedefaultendpunct}{\mcitedefaultseppunct}\relax
	\EndOfBibitem
	\bibitem[Appadurai \latin{et~al.}(2021)Appadurai, Nagesh, and
	Srivastava]{appadurai_reht_2021}
	Appadurai,~R.; Nagesh,~J.; Srivastava,~A. High resolution ensemble description
	of metamorphic and intrinsically disordered proteins using an efficient
	hybrid parallel tempering scheme. \emph{Nature Communications} \textbf{2021},
	\emph{12}, 958\relax
	\mciteBstWouldAddEndPuncttrue
	\mciteSetBstMidEndSepPunct{\mcitedefaultmidpunct}
	{\mcitedefaultendpunct}{\mcitedefaultseppunct}\relax
	\EndOfBibitem
	\bibitem[Krivov(2021)]{krivov_npne_2021}
	Krivov,~S. Non-Parametric Analysis of Non-Equilibrium Simulations.
	\textbf{2021}, arXiv: 2102.03950 [physics:chem--ph]\relax
	\mciteBstWouldAddEndPuncttrue
	\mciteSetBstMidEndSepPunct{\mcitedefaultmidpunct}
	{\mcitedefaultendpunct}{\mcitedefaultseppunct}\relax
	\EndOfBibitem
	\bibitem[Coifman and Lafon(2006)Coifman, and Lafon]{coifman_diffusion_2006}
	Coifman,~R.~R.; Lafon,~S. Diffusion maps. \emph{Appl. Comput. Harmon. Anal.}
	\textbf{2006}, \emph{21}, 5--30\relax
	\mciteBstWouldAddEndPuncttrue
	\mciteSetBstMidEndSepPunct{\mcitedefaultmidpunct}
	{\mcitedefaultendpunct}{\mcitedefaultseppunct}\relax
	\EndOfBibitem
	\bibitem[Belkin and Niyogi(2003)Belkin, and Niyogi]{belkin_laplacian_2003}
	Belkin,~M.; Niyogi,~P. Laplacian eigenmaps for dimensionality reduction and
	data representation. \emph{Neural Comput.} \textbf{2003}, \emph{15},
	1373--1396\relax
	\mciteBstWouldAddEndPuncttrue
	\mciteSetBstMidEndSepPunct{\mcitedefaultmidpunct}
	{\mcitedefaultendpunct}{\mcitedefaultseppunct}\relax
	\EndOfBibitem
	\bibitem[Tenenbaum \latin{et~al.}(2000)Tenenbaum, Silva, and
	Langford]{tenenbaum_global_2000}
	Tenenbaum,~J.~B.; Silva,~V.~d.; Langford,~J.~C. A {Global} {Geometric}
	{Framework} for {Nonlinear} {Dimensionality} {Reduction}. \emph{Science}
	\textbf{2000}, \emph{290}, 2319--2323\relax
	\mciteBstWouldAddEndPuncttrue
	\mciteSetBstMidEndSepPunct{\mcitedefaultmidpunct}
	{\mcitedefaultendpunct}{\mcitedefaultseppunct}\relax
	\EndOfBibitem
	\bibitem[Wu \latin{et~al.}(2017)Wu, Nüske, Paul, Klus, Koltai, and
	Noé]{wu_vamp_2017}
	Wu,~H.; Nüske,~F.; Paul,~F.; Klus,~S.; Koltai,~P.; Noé,~F. Variational
	Koopman models: Slow collective variables and molecular kinetics from short
	off-equilibrium simulations. \emph{J. Chem. Phys.} \textbf{2017}, \emph{146},
	154104\relax
	\mciteBstWouldAddEndPuncttrue
	\mciteSetBstMidEndSepPunct{\mcitedefaultmidpunct}
	{\mcitedefaultendpunct}{\mcitedefaultseppunct}\relax
	\EndOfBibitem
	\bibitem[Williams \latin{et~al.}(2015)Williams, Kevrekidis, and
	Rowley]{williams_data-driven_2015}
	Williams,~M.~O.; Kevrekidis,~I.~G.; Rowley,~C.~W. A {Data}-{Driven}
	{Approximation} of the {Koopman} {Operator}: {Extending} {Dynamic} {Mode}
	{Decomposition}. \emph{J. Nonlinear Sci.} \textbf{2015}, \emph{25},
	1307--1346\relax
	\mciteBstWouldAddEndPuncttrue
	\mciteSetBstMidEndSepPunct{\mcitedefaultmidpunct}
	{\mcitedefaultendpunct}{\mcitedefaultseppunct}\relax
	\EndOfBibitem
\end{mcitethebibliography}

\providecommand{\latin}[1]{#1}
\makeatletter
\providecommand{\doi}
{\begingroup\let\do\@makeother\dospecials
	\catcode`\{=1 \catcode`\}=2 \doi@aux}
\providecommand{\doi@aux}[1]{\endgroup\texttt{#1}}
\makeatother
\providecommand*\mcitethebibliography{\thebibliography}
\csname @ifundefined\endcsname{endmcitethebibliography}
{\let\endmcitethebibliography\endthebibliography}{}

\clearpage
\begin{figure}[htbp]
	\includegraphics[width=.8\linewidth]{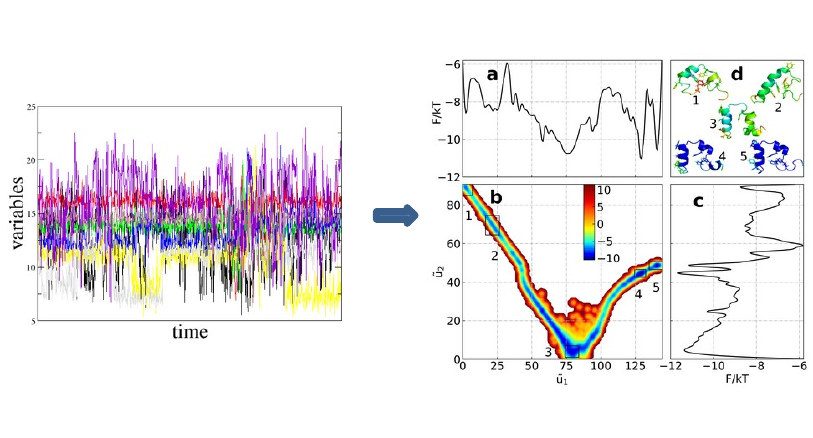}
	\captionsetup{labelformat=empty}
	\caption{For Table of Contents Only}
\end{figure}

\end{document}